\documentclass{article}

\usepackage[english]{babel}

\usepackage{iclr2023_conference,times}

\usepackage{amsmath}
\usepackage{mathtools}
\usepackage{amsfonts}
\usepackage{graphicx}
\usepackage{parskip}
\usepackage{comment}
\usepackage{csquotes}
\usepackage{geometry}
\usepackage{array}
\usepackage{booktabs}
\usepackage[most]{tcolorbox}
\usepackage{wrapfig}
\usepackage{float}
\usepackage{tabularx}
\usepackage{subcaption}
\usepackage{enumitem}
\usepackage{empheq}

\AtBeginDocument{
    \setlength{\abovedisplayskip}{1.25em}
    \setlength{\belowdisplayskip}{1.25em}
    \setlength{\abovedisplayshortskip}{0.75em}
    \setlength{\belowdisplayshortskip}{0.75em}
}

\usepackage[colorlinks=true, allcolors=blue]{hyperref}

\usepackage[colorlinks=true, allcolors=blue]{hyperref}

\def\sym#1{\ifmmode^{#1}\else\(^{#1}\)\fi}

\newcommand{\flop}{\, \textrm{FLOP}}

\newcommand{\flops}{\, \textrm{FLOP}/\textrm{s}}

\newcommand{\mac}{\, \textrm{MAC}}

\newcommand{\loss}{\mathcal{L}}

\newcommand{\oom}{\, \textrm{OOM}}

\newcommand{\mus}{\text{ }\mu\text{s}}
\newcommand{\ms}{\text{ ms}}
\newcommand{\ns}{\text{ ns}}
\newcommand{\gbs}{\, \textrm{GB}/\textrm{s}}
\newcommand{\dmodel}{{d_{\text{model}}}}
\newcommand{\dff}{{d_{\text{ff}}}}

\newcommand{\ndp}{N_{\text{DP}}}
\newcommand{\ntp}{N_{\text{TP}}}
\newcommand{\ntpmodel}{N_{\text{TP, model}}}
\newcommand{\ntpff}{N_{\text{TP, ff}}}
\newcommand{\npp}{N_{\text{PP}}}
\newcommand{\nep}{N_{\text{EP}}}
\newcommand{\ngpu}{N_{\text{GPU}}}

\newcommand{\nparams}{N_{\text{p}}}

\newcommand{\dcrit}{d'}
\newcommand{\bcrit}{b'}
\newcommand{\ncrit}{N_{\text{critical}}}

\newcommand{\tcrit}{T_{\text{critical}}}
\newcommand{\tlim}{T_{\text{limit}}}
\newcommand{\traintime}{t_\text{train}}

\newcommand{\bnetwork}{B_{\text{network}}}
\newcommand{\Bnet}{B_{\text{net}}}
\newcommand{\Bdram}{B_{\text{DRAM}}}
\newcommand{\xin}{X_\text{in}}
\newcommand{\xout}{X_\text{out}}
\newcommand{\ppp}{P_\text{PP}}
\newcommand{\pep}{P_\text{EP}}
\newcommand{\pptop}{P_\text{P2P}}
\newcommand{\questionslist}{\begin{enumerate}[label=\textbf{Q\arabic*}]
    \item Given present-day algorithms, GPUs, and interconnects, what is the biggest training run that can be performed within a fixed duration, before intra- and inter-GPU data movement starts to seriously worsen utilization or even render it impossible?
    
    \item How far might this limit be extended, and what algorithmic or hardware progress can achieve that?
\end{enumerate}}

\title{Data movement limits to frontier model training}

\author{Ege Erdil\textsuperscript{*}\\
\text{Epoch AI} \\
\texttt{ege@epochai.org}
\And
David Schneider-Joseph\textsuperscript{*}\\
\texttt{david@davidsj.com}}
\date{\today}

\iclrfinalcopy

\begin{document}
\def\thefootnote{*}\footnotetext{Equal contribution.}
\def\thefootnote{\arabic{footnote}}

\begin{center}
\maketitle
\end{center}

\begin{abstract} 
We present a theoretical model of distributed training, and use it to analyze how far dense and sparse training runs can be scaled. Under our baseline assumptions, given a three month training duration, data movement bottlenecks begin to significantly lower hardware utilization for training runs exceeding about $10^{28}$ FLOP, two orders of magnitude above the largest training run to date, \textbf{suggesting the arrival of fundamental barriers to scaling in three years} given recent rates of growth. A training run exceeding about \( 10^{31} \flop \) is infeasible even at low utilization. However, more aggressive batch size scaling and/or shorter and fatter model shapes, if achievable, have the potential to permit much larger training runs. An interactive version of our model will shortly be accessible \href{https://epochai.org/tools/distributed-training}{here}.
\end{abstract}

\section{Introduction}
\label{sec:introduction}

Scaling up neural networks and training them on more examples is crucial for good task performance \citep{hestness2017scaling, kaplan2020scaling,hoffmann2022training, deepseekai2024deepseek}, with state-of-the-art models requiring tens of thousands of GPUs\footnote{We focus on GPUs, but our theoretical model and findings are broadly applicable to other accelerators, and even groups of accelerators.} to train in a reasonable duration. Previous work \citep{huang2019gpipe, rajbhandari2020zero, lepikhin2020gshard, narayanan2021efficient, jiang2024megascale} has developed practical techniques enabling the rapid scaling of the past decade \citep{sevilla2022compute}.

In this work, we address unexamined fundamental questions about \textbf{limits to scaling in the future:}

\questionslist

Answering these questions empirically would require millions of GPUs and large-scale engineering efforts, so we instead approach them theoretically. In doing so, we develop a simulator that can find optimal training run configurations accounting for the factors that we identify as fundamental. This gives us the answers:

\begin{enumerate}[label=\textbf{A\arabic*}]
    \item With most current technology, \textbf{GPU utilization starts to fall at \( \mathbf{\approx 10^{28}}\) floating point operations (FLOP), about three years away at recent trends} \citep{EpochNotableModels2024} of $4.2\times$ growth per year.

    Currently-available specialized high-bandwidth inter-node interconnects can permit training runs about two orders of magnitude larger ($\approx 10^{30} \flop$), at which point latencies begin to worsen utilization, until reaching \textbf{an absolute latency barrier at $\approx 10^{31} \flop$, about seven years away.}
    
    \item Improved hardware interconnects may buy no more than two orders of magnitude in training run size, assuming technology anything like the current paradigm. Beyond that, the critical innovations must come from machine learning algorithms: \textbf{The key challenge is transforming two serial dependencies — between batches and between layers — into opportunities for parallelism, by making batch sizes bigger (perhaps enabled by sparsity) and models wider and shallower.} Achieving these goals may be quite difficult in practice.
\end{enumerate}

\begin{figure}[htbp]
    \centering
    \begin{subfigure}[b]{0.49\textwidth}
        \centering
        \begin{minipage}[b]{0.89\linewidth}
            \centering
            \caption*{\centering \textbf{Data movement and latency bottlenecks limit the scale of training runs to $\boldmath{10^{28} \text{ to } 10^{31}}$ FLOP}}
        \end{minipage}
        \includegraphics[width=\textwidth]{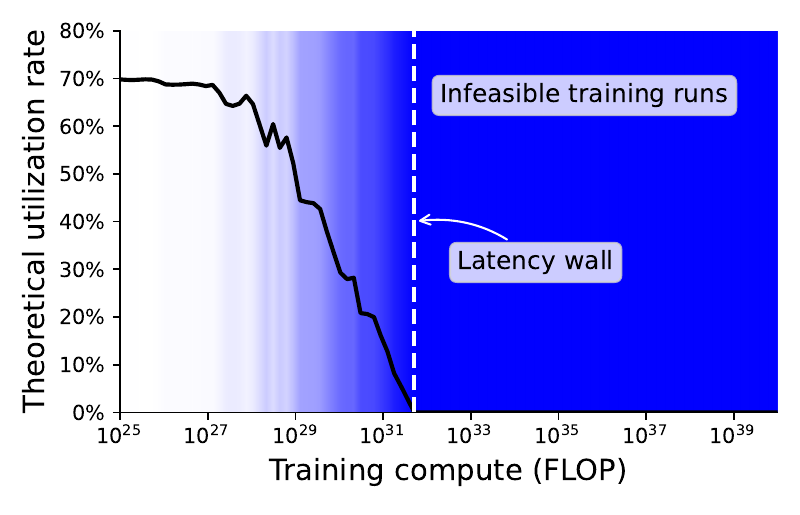}
        \label{fig:fig1}
    \end{subfigure}
    \hfill
    \begin{subfigure}[b]{0.49\textwidth}
        \centering
        \begin{minipage}[b]{0.89\linewidth}
            \centering
            \caption*{\centering \textbf{New scaling techniques and hardware innovations could push back limits to scaling}}
        \end{minipage}
        \includegraphics[width=\textwidth]{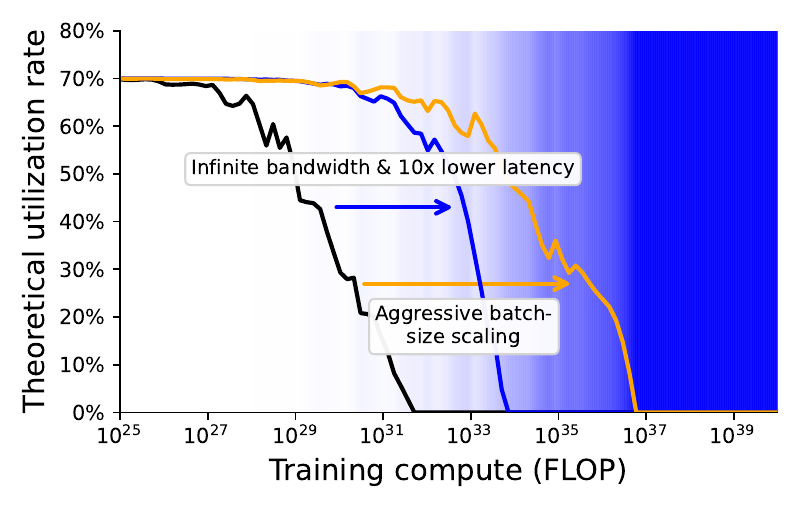}
        \label{fig:fig2}
    \end{subfigure}
    \vspace{-2em}
    \caption{With current technology, such as the H100 GPU and current scaling techniques, data movement bottlenecks lower hardware utilization for training runs exceeding $10^{28}$ FLOP, and a ``latency wall" renders surpassing $10^{31}$ FLOP infeasible (left). However, with innovations in scaling (such as techniques to enable much larger batch sizes) or dramatic increases in network bandwidth coupled with a 10$\times$ reduction in inter- and intra-GPU latency, training runs can be at least a few orders of magnitude larger (right).}
    \label{fig:side_by_side}
\end{figure}

This work is organized as follows:

\begin{itemize}
    \item Section \ref{sec:toymodel} introduces a simplified model of a neural network consisting of stacked sparse linear multi-layer perceptrons that we use as the basis for our analysis throughout the paper.

    \item Section \ref{sec:parallelism} provides an overview of the four main parallelism strategies employed in distributed training—data, tensor, pipeline, and expert parallelism—and summarizes their communication costs.

    \item Section \ref{sec:constraints} identifies the key factors that constrain distributed training, including data movement, critical batch size, and latency, and model depth.

    \item Section \ref{sec:closed-form-biggest-training-run} derives closed-form expressions for the maximum training scale under this model.

    \item Section \ref{sec:complete-model} presents the complete theoretical model accounting for all identified constraints and discusses simulation results on current hardware, demonstrating the limits to efficient scaling.
\end{itemize} 

We \hyperref[sec:conclusion]{conclude} by discussing the implications of our findings, and posing key open technical questions whose answers will determine the limits to large-scale model training.

\section{A toy model: stacked sparse linear MLP blocks}
\label{sec:toymodel}
We first carefully define the class of models to be considered. Since the Transformer \citep{vaswani2017attention}, including its sparse varieties \citep{fedus2022switchtransformersscalingtrillion}, is the dominant architecture today for frontier models, it seems a natural baseline. As we will show momentarily, the great majority of computation when training a Transformer occurs in its linear layers, so we adopt a simplified model consisting of these elements. This approach has the advantage that such layers also constitute the bulk of computation in many alternative architectures (\cite{peng2023rwkv} and \cite{gu2023mamba} among others), so our model retains applicability to a wide variety of potential future algorithmic developments. However, the huge space of possible unexpected algorithmic developments precludes any watertight guarantees, should state-of-the-art architectures or learning algorithms change drastically.

The formal definitions we make are as follows: a \textbf{sparse linear multi-layer perceptron (MLP) block} (``MLP block" for short) consists of learnable weight matrices \( W_1^e \in \mathbb R^{\dff \times \dmodel} \) and \( W_2^e \in \mathbb R^{\dmodel \times \dff} \) for \textbf{expert indices} \( e \) in the range \( 1 \leq e \leq E \), where \( E \) is the \textbf{sparsity factor} of the model.

Given a matrix of input vectors \( \xin \in \mathbb R^{\dmodel \times b} \), where \( b \) is the batch size in units of tokens, a router $\rho:~\mathbb R^{\dmodel} \to \{1, \ldots, E\}$ (assumed to be computationally inexpensive) chooses some expert index \( \rho(\xin^k) \) for each token index \( k \le b \). The sparse linear MLP $f$ then yields an output matrix \( \xout = f(\xin) \in \mathbb R^{\dmodel \times b}\) by mapping each token's input column \( \xin^k \) to the corresponding output column \( \xout^k \) via the weight matrices corresponding to the chosen expert. If $^{(e)}$ indexes the set of tokens routed to expert $e$, then:
\[ \xout^{(e)} = W_2^e \Big(W_1^e \xin^{(e)}\Big). \]
We optimistically assume \textit{balanced routing} (Section \ref{sec:expert-parallelism}): the same number $b/E$ of tokens is routed to each of the $E$ experts, so that during the forward pass every expert performs matrix multiplications of shapes $(\dff, \dmodel) \times (\dmodel, b/E) \to (\dff, b/E)$ and then $(\dmodel, \dff) \times (\dff, b/E) \to (\dmodel, b/E)$, each requiring \( \dmodel \dff b/E \) multiply-accumulate (MAC) operations. Across both linear layers of all $E$ experts, the MLP block's total arithmetic cost is \( 2 \dmodel \dff b \) MAC. We then assume \( L \) MLP blocks in total, so that the model's forward pass is their composition:
\[ F(X) = f_L(f_{L-1} \cdots (f_1(X))). \]
Across these \( L \) blocks, the model has
\begin{align}
    \label{eq:nparams}
    \nparams = 2 LE \dmodel \dff
\end{align}
parameters in total and a forward pass on a batch size of \( b \) requires \(\nparams b/E = 2L \dmodel \dff b \) MAC.

During the backward pass, gradients of the loss $\loss$ must be computed for the inputs (activations and weights) of each matrix multiplication. A matrix multiplication $C = AB$ of shape $(I, K) \times (K, J) \to (I, J)$ on the forward pass becomes two matrix multiplications,
\begin{align*}
    \partial \loss / \partial A =&\ (\partial \loss / \partial C)B^\top,\\
    \partial \loss / \partial B =&\ A^\top (\partial \loss / \partial C),
\end{align*}
of shapes $(I, J) \times (J, K) \to (I, K)$ and $(K, I) \times (I, J) \to (K, J)$ on the backward pass, all requiring the same number $IJK$ of MAC. Thus accounting for the backward pass, the number of MAC for our model $F$ is tripled and becomes $6L \dmodel \dff b$.

An actual Transformer has linear layers not only in its MLP but also for attention queries, keys, values, and outputs. These can be fused with and computed in parallel with the MLP block \citep{wang2021gpt, chowdhery2022palm}, so that our simplified model encompasses these. However, there are also operations which we neglect:

\begin{itemize}
    \item Element-wise operations such as nonlinear activation functions,\footnote{Logically, dropping the nonlinearities makes a dense Transformer architecture trivial, as the end-to-end map \( F \) becomes linear and thus representable by a single matrix. We assume the nonlinearities are still present, but omit modeling them due to their relative computational insignificance.} layer normalization \citep{ba2016layernormalization}, and residual accumulation \citep{he2015deepresiduallearningimage}. We treat these as negligible as they involve $O(\dmodel)$ or $O(\dff)$ computations per token, as compared to the linear layers' $O(\dmodel \dff)$ computations per token. For large language models, \( \dmodel \) is typically on the order of \( 10^4 \), and \( \dff \) a multiple thereof, so relative negligibility tends to hold in practice. Furthermore, these element-wise operations need not impose additional significant data movement of their own, as they can often be fused with matrix multiplication kernels.

    \item Embedding and unembedding projections $W_\text{embed}~\in~\mathbb R^{\dmodel \times V}, W_\text{unembed}~\in~\mathbb R^{V \times \dmodel}$. As these are computed only once per forward pass rather than once per Transformer block, and the vocabulary size $V$ is typically not much larger than $\dff$, we ignore these.

    \item Scaled dot-product attention scoring and the corresponding linear combination of the attention head value vectors. For a sequence length $\ell$ and typical attention width $n_\text{heads} \cdot d_\text{head} \approx \dmodel$, this requires $O(\ell \dmodel b)$ MAC per attention layer each batch. While not completely negligible, this cost is typically small compared to the MLP MAC \( \approx 2 \dmodel \dff b \) above, as the sequence length $\ell$ is usually significantly smaller than the internal MLP width $\dff$ during most of training. \citep{dubey2024llama3herdmodels, devlin2019bertpretrainingdeepbidirectional}
\end{itemize}

We thus see that our toy model, despite being dramatically simpler to analyze and understand than an actual Transformer, encompasses most of its arithmetic work.

We have so far considered only arithmetic costs. To understand bottlenecks to distributed training, we must also investigate the costs of data movement, both within and between GPUs. Much of what follows will concern itself with this investigation.\footnote{Beyond our toy model, element-wise operations need not impose additional significant data movement of their own, as they can often be fused with matrix multiplication kernels. Even when not, the volume tends to be small relative to weight matrices due to slicing along the batch dimension from data and pipeline parallelism into very small nanobatches (Section \ref{sec:utilization-cliff}), necessitating many accesses of the same weight matrix per batch. Scaled dot-product attention can impose additional data movement, especially if the nanobatch size is significantly smaller than the sequence length, which will be one of several factors causing our results to err on the optimistic side. Actual achieved training run sizes and utilizations will likely be bounded above by our model.} To proceed, we must first understand how the work of a training run can be distributed across GPUs.

\section{Methods of Parallelism}
\label{sec:parallelism}

The workload of a training run can be distributed through a variety of methods \citep{openai2022largemodels}: data parallelism (DP), tensor parallelism (TP), pipeline parallelism (PP), and expert parallelism (EP), corresponding to the four problem dimensions: batch size, layer width, model depth, and sparsity, respectively.\footnote{For a Transformer, \textit{sequence parallelism} is also available, partitioning work even across different tokens in the same sequence. This can be thought of as data parallelism with some additional communication for the attention layers. We omit discussion of this method because of our restricted focus on linear layers.} This list of parallelism methods is also exhaustive, because a single matrix multiplication only allows for tensor and data parallelism. Further parallelism must come from assigning different matrix multiplications to different GPUs, which can be done vertically (pipeline parallelism) or horizontally (expert parallelism). In this section, we briefly overview these methods, organized by their characteristic communication patterns.

\subsection{Tensor slicing and all-reduce: data and tensor parallelism}

Consider a forward pass matrix multiplication $C = AB$ of shape $(I, K) \times (K, J) \to (I, J)$ and its backward pass computations $\partial \loss / \partial A = (\partial \loss / \partial C)B^\top$ and $\partial \loss / \partial B = A^\top (\partial \loss / \partial C)$ as described in Section \ref{sec:toymodel}. For example, $A$ could be the weight matrix $W_1^e$ for an expert's first layer, and $B$ its input activations $\xin^{(e)}$, in which case the forward pass multiplication shape is $(\dff, \dmodel) \times (\dmodel, b/E)$, as discussed earlier.

\subsubsection{One-dimensional slicing}

A natural idea is to partition the data (and associated work) along one of the dimensions of size $I$, $J$, or $K$, while replicating the other dimensions. When this is the expert's batch dimension of size $b/E$, we call this \textbf{data parallelism,} and when it's one of the layer shape dimensions $\dmodel$ or $\dff$, we call it \textbf{tensor parallelism.} Fundamentally they require the same communication pattern, which we now describe in general terms.

If the data is partitioned along the ``internal" $K$-sized dimension into $K'$-sized chunks across $N_K = K/K'$ GPUs, and the computation partitioned accordingly, then the $n^\text{th}$ GPU performs a matrix multiplication of shape $(I, K') \times (K', J) \to (I, J)$, defined by:
\[c_{ij}^n = \sum_{\mathclap{k=nK'}}^{\mathclap{(n+1)K'-1}} a_{ik}b_{kj}.\]
Each GPU thus has only a partially-reduced value for each component of $C$, but requires the fully-reduced component $c_{ij} = c_{ij}^0 + c_{ij}^1 + \ldots + c_{ij}^{N_K-1}$ so that it may proceed with the next computation step. This necessitates an \textbf{all-reduce} collective communication across the $N_K$ GPUs. \citep{kumar1994introduction}

We can derive the minimal inter-GPU data movement volume for this all-reduce, assuming that messages contain the single-word partial sum accumulated so far by the transmitting GPU. Given any total ordering of messages consistent with their causal partial ordering, the $(N_K - 1)^\text{th}$ message is the first whose receiver can possibly lie causally downstream of all other GPUs and therefore contain the fully-reduced component $c_{ij}$. The other $N_K - 1$ GPUs must then also receive at least one additional message, so there must be at least $2(N_K - 1)$ messages, and hence words, received in total.\footnote{Strictly speaking, not every word need be received by a \textit{GPU} per se, but merely by some network device. For example, the network fabric can cut GPU all-reduce bandwidth requirements by about $2\times$. \citep{graham2016scalable}}

That is, to all-reduce a single word across $N_K$ GPUs, each GPU must receive
\begin{align}
    \label{eq:allreduce-words}
    2(N_K - 1)/N_K \approx 2
\end{align}
words.

There are $I \cdot J$ such components, so this matrix multiplication's all-reduce must cost at least
\[2 IJ (N_K - 1)\]
words of inter-GPU data movement.\footnote{Here and throughout, we count only words received. Words transmitted will always either be equal, or (in a multicast setting) at least proportional.} This lower bound can in fact be achieved exactly by, for example, consecutively performing a reduce-scatter and an all-gather operation \citep{kumar1994introduction}.

On the backward pass matrix multiplications of shapes $(I, J) \times (J, K) \to (I, K)$ and $(K, I) \times (I, J) \to (K, J)$, the $K$-sized dimension is ``external", hence requires no reduction, and therefore no inter-GPU communication.

If instead the data is partitioned across $N_I$ GPUs along the $I$-sized dimension, then this dimension is ``internal" (i.e. requiring reduction) only for the second backward pass matrix multiplication of shape $(K, I) \times (I, J) \to (K, J)$, imposing
\[2 KJ (N_I - 1)\]
words of inter-GPU data movement in the backward pass, and none in the forward pass. Symmetrically, a partition across $N_J$ GPUs along the $J$-sized dimension requires
\[2 IK (N_J - 1)\]
words of inter-GPU data movement in the backward pass, and none in the forward pass.

\subsubsection{Multi-dimensional slicing}
\label{sec:multidim-slicing}

In general, we may partition the data across $N = N_I N_J N_K$ GPUs, with each performing a forward pass multiplication of shape $(I', K') \times (K', J') \to (I', J')$, where as before we define $I' = I/N_I$, etc.

Because communication in the forward pass happens only along the ``inner" $K$-sized dimension, this can be treated as $N_I N_J$ independent matrix multiplications, entailing (as derived above) $2I'J'(N_K - 1)$ words of inter-GPU data movement each. Across all independent multiplications, this works out to $2(N_I I') (N_J J')(N_K - 1) = 2IJ(N_K - 1)$ words of inter-GPU data movement in the forward pass, exactly as in the $K$-only parallelism case.

We find similarly for backward pass communication across the $I$ and $J$ dimensions, so that:
\begin{align}\label{eq:allreduce-data-mvmt}
\text{total inter-GPU data movement}\ &= 2[IJ(N_K - 1)
+ KJ(N_I - 1)
+ IK(N_J - 1)] \text{ words.}
\end{align}
Let us now consider intra-GPU data movement between main DRAM memory and the logic chip.\footnote{See Appendix \ref{sec:appendix-model-description} for a discussion of data movement between SRAM-based caches, shared memory, register banks, and execution units.} Assuming an ideal cache, each forward pass matrix multiplication loads $I'K' + K'J'$ words from memory and writes $I'J'$ words to memory, for
\[I'K' + K'J' + I'J'\]
words accessed per GPU in the forward pass. The two backward pass multiplications behave similarly, tripling total memory IO, so that across all $N_I N_J N_K$ GPUs we have\footnote{In fact, caching is imperfect, increasing required IO, while some opportunities for cache hits across matrix multiplications can decrease this IO. We treat this as a first-order approximation in the regime where the matrices on each GPU are large enough that they would not fit into SRAM -- our full model also considers the scenario where the matrices are sufficiently small that we can cut out DRAM reads and writes altogether.}
\begin{align*}
\text{total intra-GPU data movement} = 3(IJN_K
+ KJN_I
+ IKN_J) \text{ words}.
\end{align*}
Note the similarity to the inter-GPU data movement formula Eq. \ref{eq:allreduce-data-mvmt} above, both having $N_I, N_J, N_K$-dependent term proportional to $IJN_K + KJN_I + IKN_J = IJK(1/I' + 1/J' + 1/K')$, or simply
\begin{equation}
    \label{eq:data-mvmt-objective}
    1/I' + 1/J' + 1/K',
\end{equation}
when we hold the problem dimensions $I, J, K$ constant.

If we also hold the total number of GPUs $N$ -- and thus the per-GPU problem size $I'J'K' = IJK/N$ -- constant, we find Eq. \ref{eq:data-mvmt-objective} is minimized when $I' = J' = K'$, that is when all three degrees of parallelism are sized so as to give each GPU a cube-shaped work unit. In practice the constraint that $N_I, N_J, N_K$ be positive integers means this exact condition is rarely achieved, and when $N$ is particularly small, the smallest dimensions go un-parallelized. As our concern is the limits to distributed training, we typically consider regimes where $N$ is quite large and the condition approximately holds.

We now apply this analysis to the concrete cases of data and tensor parallelism.

\subsubsection{Data parallelism}

In this case, each expert's activations $\xin^{(e)}$ and $\xout^{(e)}$ (as well as intermediate states $W_1^e\xin^{(e)}$) are sliced $\ndp$-way along the $b/E$-sized batch dimension. This becomes a reduction dimension in the backward pass when computing gradients with respect to the (replicated) weight matrices $W_1^e$ and $W_2^e$ (via multiplications of shapes $(\dff, b/E) \times (b/E, \dmodel)$ and $(\dmodel, b/E) \times (b/E, \dff)$), so \textit{data parallelism necessitates an all-reduce of weight gradients across the batch dimension}, with inter-GPU data movement per weight matrix of $2\dff \dmodel(\ndp - 1)$ words per batch, as can be seen by applying Eq. \ref{eq:allreduce-data-mvmt} to this case.

 \begin{figure}[h!]
    \centering
    \includegraphics[width=1\textwidth]{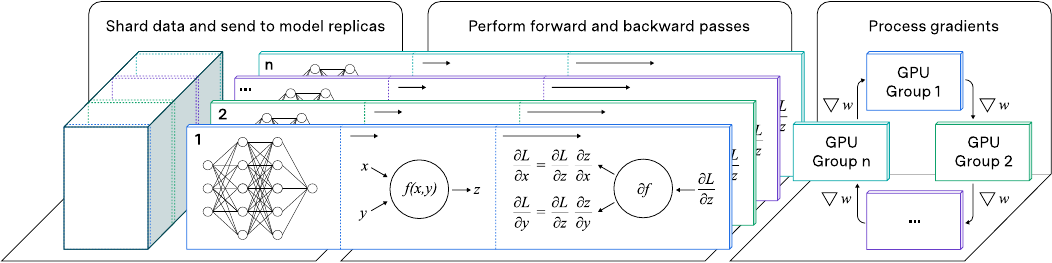}
\caption{\small Data parallelism. The input data is divided into shards and processed independently by multiple model replicas. Each replica computes gradients for its local shard, which are then all-reduced across all replicas to obtain the full batch gradient. This full gradient finally updates the model parameters on each replica.}

    \label{fig:dp-visual}
\end{figure}

Aggregating across the $L$ MLP blocks, $E$ experts per block, and two layers per expert, this amounts to a total inter-GPU data movement of $4LE\dff \dmodel(\ndp - 1)$ words per batch, i.e.
\[2\nparams(\ndp - 1)/\ndp\]
words per batch per data-parallel worker, where $\nparams = 2LE\dff\dmodel$ is the total number of model parameters.

Across those $E$ experts, each data-parallel worker sees $b/\ndp$ tokens. Naively implemented, each stores a full copy of the $\nparams$ model parameters and associated optimizer state. However, the optimizer state can be sharded by inserting the optimizer step \textit{in between} a reduce-scatter/all-gather all-reduce implementation: first the gradients are reduce-scattered, then each worker optimizes the weights corresponding to its gradient partition, then the new weights are all-gathered across workers. \citep{rajbhandari2020zero}

\subsubsection{Tensor parallelism}
\label{sec:tensor-parallelism}

In this case, an expert's activations and weights are sliced into $\ntp = \ntpff \times \ntpmodel$ partitions, along the $\dff$ and $\dmodel$ dimensions. The $\dmodel$ dimension is internal in the first layer's forward pass of shape $(\dff, \dmodel) \times (\dmodel, b/E)$, as well as the second layer's activation gradient computation of shape $(\dff, \dmodel) \times (\dmodel, b/E)$. Symmetrically, the $\dff$ dimension is internal in the second layer's forward pass and the first layer's activation gradient computation. Applying Eq. \ref{eq:allreduce-data-mvmt}, we see $2(b/E)[\dff (\ntpmodel - 1) + \dmodel (\ntpff - 1)]$ words of inter-GPU data movement for each of an expert's two layers. Aggregating across the $L$ MLP blocks, $E$ experts per block, and two layers per expert, we have total inter-GPU data movement of
\[4Lb[\dff (\ntpmodel - 1) + \dmodel (\ntpff - 1)] / \ntp\]
words per batch per tensor-parallel worker.

\begin{figure}[h!]
    \centering
    \includegraphics[width=1\textwidth]{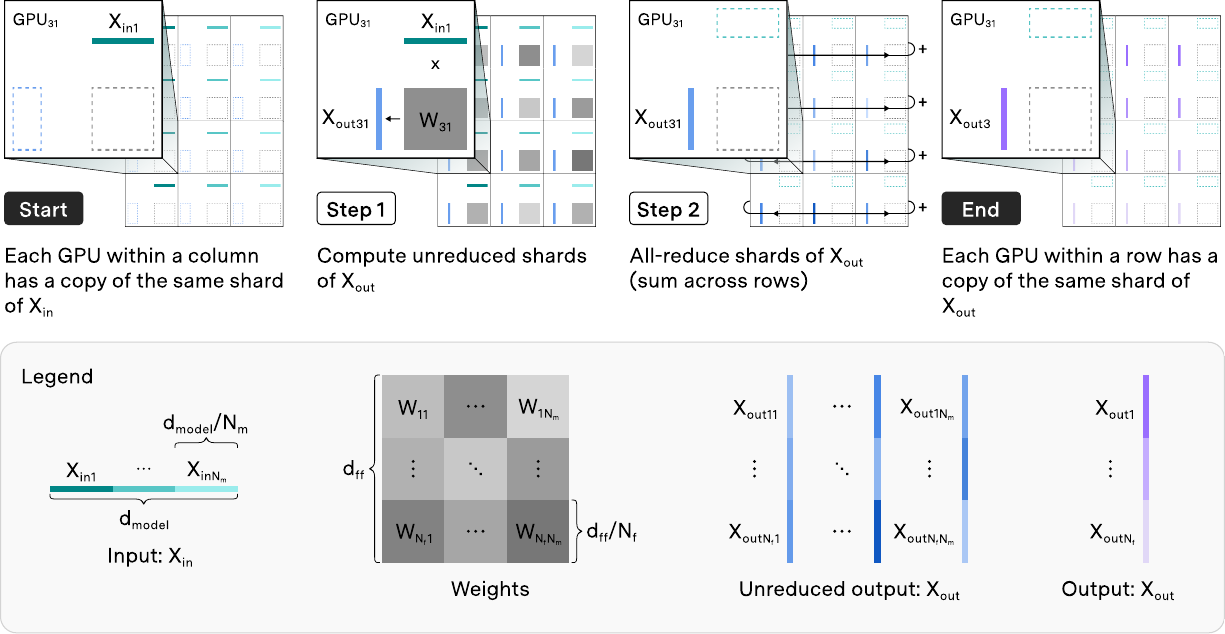}
\caption{\small 2D tensor parallelism for an MLP block expert's first layer weight matrix $W$. Start: Input vector $X_{\text{in}}$ is scattered ``horizontally'' (across $N_m = \ntpmodel$ GPUs) and duplicated ``vertically" (across $N_f = \ntpff$ GPUs), and weight matrix shards $W_{11}$ through $W_{N_fN_m}$ are scattered along $d_{\text{ff}}$ (rows) and $d_{\text{model}}$ (columns) in the same set of GPUs. Step 1: Each GPU computes a partially-reduced shard of output matrix $\xout$. Step 2: All-reduce operation across GPU rows computes fully-reduced shards of $\xout$. End: Each GPU on a given row has an identical, fully-reduced copy of the $\xout$ shard corresponding to that row.}
\label{fig:tp-visual}
\vspace{-10pt}
\end{figure}

From our general analysis above, we see that when $\ntp$ is small (the norm today), the smaller dimension $\dmodel$ goes un-partitioned ($\ntpmodel = 1$), yielding \textbf{1D tensor parallelism}, but that otherwise (as for the much larger training runs which are our focus), \textbf{2D tensor parallelism} becomes optimal, with $\dff/\ntpff \approx \dmodel/\ntpmodel$ (i.e. roughly square weight partitions). Solving this yields approximately
\[8Lb\sqrt {\dmodel \dff / \ntp}\]
words of inter-GPU data movement per batch per tensor-parallel worker.

\subsection{Point-to-point: pipeline and expert parallelism}

We have so far examined partitioning the work along the per-expert batch dimension of size $b/E$ (data parallelism) and the two weight matrix dimensions of size $\dmodel$ and $\dff$ (tensor parallelism). Two problem dimensions remain: we can also partition depth-wise across the $L$ MLP blocks with \textbf{pipeline parallelism,} and/or across the $E$ experts with \textbf{expert parallelism.} In this case, a given token at a given layer is processed by the GPU (or GPUs, if combined with tensor parallelism) corresponding to that layer's \textbf{pipeline stage} and that token's routed expert. As the token moves through the layers during the forward and backward passes, its activations are transferred in a simple point-to-point communication of $\dmodel$ words whenever crossing pipeline- \textit{or} expert-parallel ranks.\footnote{To avoid redundant communication due to replication across tensor-parallel ranks, tensor-parallel all-reduces can be split into two phases: a reduce-scatter \textit{before} the point-to-point communication, followed by an all-gather on the new set of tensor-parallel peers \textit{after} the point-to-point communication. \citep{narayanan2021efficient}} Because of the disjunctive nature of this condition, the point-to-point communication cannot always be attributed solely to one or the other method of parallelism, and a joint analysis is warranted.

Indeed, when the expert-parallel degree $\nep$ is even moderately large and (as we shall assume) expert routing is stochastically independent of token and layer, then expert-parallel communication \textit{usually} necessitates a point-to-point communication after each MLP block, and so pipeline-parallel communication comes along nearly ``for free". However, pipeline parallelism comes with its own challenges, which we now examine.

\subsubsection{Pipeline parallelism}
\label{sec:pipeline-parallelism}

In a naive pipeline-parallel configuration, the first pipeline stage is assigned the first \( L/\npp \) layers, the second is assigned the next \( L/\npp \) layers, \textit{et cetera}. Consequently, in each forward and backward pass, a vector of size \( \dmodel \) has to be communicated a total of \( \npp - 1 \) times. The total data movement cost across all GPUs is simply \( \dmodel (\npp - 1) \) activations per forward or backward pass per token.

\begin{figure}[t!]
    \centering
    \includegraphics[width=1\textwidth]{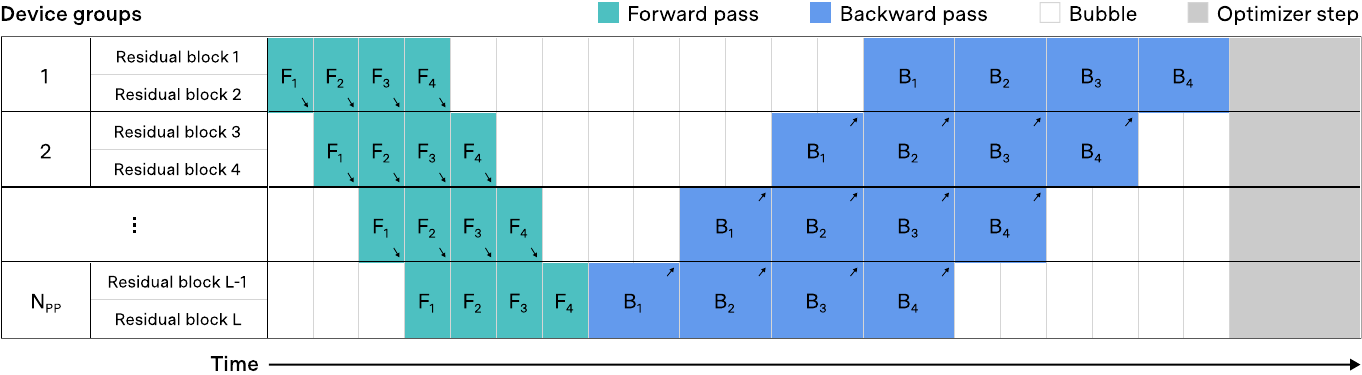}
\caption{\small Pipeline parallelism. This diagram depicts the sequential execution flow of microbatches \(F_1\) to \(F_4\) during the training of a deep learning model using a GPipe pipeline schedule \citep{huang2019gpipe}. Device groups, each containing one or more GPUs, are allocated distinct sets of model layers, shown here as residual blocks, and they process the microbatches in a staggered fashion. Pipeline bubbles, the white areas, indicate periods of GPU inactivity due to dependency waits. The optimizer step, highlighted in gray, follows the backward passes and is where model parameter updates occur.}
    \label{fig:pp-visual}
\vspace{-10pt}
\end{figure}

Because model layers are inherently serial, pipeline parallelism imposes a unique challenge not encountered by the other forms -- data, tensor, and expert -- of parallelism: minimizing the so-called \textbf{pipeline bubble}, the time GPUs spend idle waiting for activations from other pipeline stages. If the entire batch were sent through the pipeline all together, then at any given moment only one of the $\npp$ pipeline stages would ever be active, and the \textbf{bubble fraction} (proportion of time spent idle) would equal $1 - 1/\npp$. So instead, the batch must be sliced further than whatever is imposed by data parallelism alone, into smaller pieces called \textbf{microbatches,} which travel individually through the pipeline according to a \textbf{pipeline schedule} in which, ideally, the large majority of pipeline stages are active most of the time.
\cite{narayanan2021efficient} notes that a naive pipelining strategy (Fig. \ref{fig:pp-visual}) with \( m \) microbatches and \( \npp \)-way pipeline parallelism incurs a bubble fraction of
\begin{equation}
    \label{eq:naive-pp-bubble-fraction}
    B_{\text{pp}} = \frac{\npp - 1}{\npp - 1 + m},
\end{equation}
or a maximum arithmetic utilization of 
\[ U_{\text{pp}} = 1 - B_{\text{pp}} = \frac{m}{\npp - 1 + m}. \]
This can be seen most easily from the perspective of the last stage: on the forward pass it spends $\npp - 1$ bubble steps waiting for that same number of previous stages, then has $m$ steps of work to perform; the backward pass looks the same in reverse. Since all stages have the same amount of work to do, they must also have this same bubble fraction.

Therefore, it's crucial to ensure that the number of microbatches $m$ is large relative to $\npp - 1$, the number of inter-stage interfaces in the pipeline. But this lowers the arithmetic intensity of the computations done by the GPUs by slicing the data matrix further along the per-expert batch dimension \( b/E \), increasing data movement costs \textit{internal to the GPUs,} as the weights and their accumulated gradients must be accessed \textit{once for each microbatch.} Increasing the batch size \( b \) is a natural answer to this conundrum, but this too has limits, which we discuss in Section \ref{sec:critical-batch-size}.

\textbf{Pipeline interleaving.}\label{sec:pipeline-interleaving} Another solution described by \cite{narayanan2021efficient} is \textit{pipeline interleaving.} By assigning non-adjacent layers of our model to the same pipeline stage, each microbatch travels through the pipeline \textit{multiple times,} for a different set of model layers each time. This increases the effective microbatch count, shortens the initial (and terminal) serial wait time for work to reach (and clear) all stages, and thereby reduces the bubble fraction \( B_{\text{pp}} \). This comes at the expense of increased network communication, but this trade-off is usually quite favorable, as the point-to-point communication volume imposed by pipeline parallelism (and shared with expert parallelism) is not large.

We leave a detailed discussion of pipeline interleaving to Appendix \ref{sec:appendix-interleaving}. Here, we note only that if the number of microbatches \( m \) is at least equal to the number of pipeline stages \( \npp \), we can pick any divisor \( i \) of \( L/\npp \) and change the pipeline schedule to achieve the following two outcomes:

\begin{enumerate}
    \item The bubble fraction becomes \( B_{\text{interleaved PP}}(i) = (\npp - 1)/((\npp-1) + im) \): the effective number of microbatches determining the bubble fraction goes up by a factor of \( i \).
    \item The total communication cost per forward pass becomes \( \dmodel (\npp \cdot i - 1) \): the effective number of pipeline stages determining network communication costs goes up by this same factor \( i \).
\end{enumerate}

\textbf{Zero bubble pipeline parallelism.}\label{sec:zero-bubble-pp} \cite{qi2023zero} recently observed that only \textit{activation} gradients are serial dependencies in the backward pass for previous pipeline stages, and hence the computation of \textit{weight} gradients can be deferred to otherwise idle time in order that those earlier pipeline stages may sooner have work. In this way, their ZB-H2 schedule completely eliminates bubbles. Though this schedule has approximately twice the memory footprint of the interleaved schedule from the previous section, this is a negligible cost in large clusters in which device memory is abundant.

The ZB-H2 schedule requires at least \( 2 \npp - 1 \) microbatches to achieve the zero bubble condition, but this is small compared to the much larger number of microbatches required even to approximate this in a traditional schedule.

\subsubsection{Expert parallelism}
\label{sec:expert-parallelism}

Expert parallelism involves slicing each layer along the expert dimension \( E \): each pair of weight matrices \( W_1^j \) and \( W_2^j \) for \( 1 \leq j \leq E \) is stored on one of $\nep$ expert-parallel ranks, with activation vectors routed to the appropriate rank for their routed expert. Even without any expert parallelism, the sparsity factor $E$ itself already harms the arithmetic intensity of matrix multiplications by shrinking the per-multiplication batch size, and expert parallelism does not in general make matters worse. However, scaling this sparsity factor could potentially increase the largest batch size which can efficiently be used in training (Section \ref{sec:critical-batch-size}

In our toy model (Section \ref{sec:toymodel}), we consider only settings where each activation vector is routed to a single expert, in which case the point-to-point communication pattern is low-bandwidth and can coincide with that already needed for pipeline parallelism. If tokens are instead routed to multiple experts \citep{shazeer2017outrageouslylargeneuralnetworks, patel2023gpt4}, with a linear combination taken of their outputs, then a tensor-parallel-style all-reduce across experts is required, potentially across unpredictable levels of the network hierarchy depending on expert placement. As our objective in this paper is to shed light on the limits to distributed training, we make optimistic assumptions when possible. In line with this, we treat the ``fine-grained" case as an extension of tensor parallelism and do not model it separately, simplifying our analysis greatly.

Continuing in this spirit, we assume \textit{balanced routing} between experts, independent of token or layer, to ensure uniform workloads across workers. \citep{lepikhin2020gshard, lewis2021baselayers, zhou2022expertchoice}

\subsection{5D parallelism}
\label{sec:5d-parallelism}

\textit{5D parallelism} involves combining all of the parallelism methods we've discussed so far: data, tensor (both dimensions), pipeline, and expert parallelism. Table \ref{tab:parallelism-summary} summarizes the network communication costs of these methods on an equal footing: how much communication they require per gradient step taken.

\renewcommand{\arraystretch}{1.5}

\begin{table}[h]
\centering
\begin{tabular}{l|l|l|l}
                     &  \vtop{\hbox{\strut Network bandwidth}\hbox{\strut per gradient step}}    & Slices along...      & \vtop{\hbox{\strut Communications can}\hbox{\strut coincide with...}}  \\ \hline
Data parallelism     & \( 2 \nparams (\ndp - 1) \)                                               & \( b \)              & Nothing \\ \hline
Tensor parallelism   & \( \approx 8bL \sqrt{\dff \dmodel \ntp} \) (for large $ \ntp $)           & \( \dmodel, \dff \)  & Nothing \\ \hline
Pipeline parallelism & \( 2b \dmodel (\npp \cdot i - 1) \)                                       & \( b, L \)           & Expert parallelism \\ \hline
Expert parallelism   & \( 2b \dmodel (L - 1) \cdot (\nep - 1)/\nep \)                                  & \( E \)            & Pipeline parallelism
\end{tabular}
\caption{The different parallelism methods available along with their network bandwidth costs. Here, $\nparams$ is the number of parameters, $b$ the batch size (in tokens), $L$ the number of MLP blocks, $\dmodel$ and $\dff$ the model widths, and \( i \) the pipeline interleaving factor.}
\label{tab:parallelism-summary}
\end{table}

To see why combining multiple parallelism methods is more efficient than using any one method of parallelism by itself, consider the isolated problem of minimizing the total network bandwidth cost for a cluster given a fixed cluster size \( \ngpu = \ntp \npp \ndp \) for a dense training run (so that \( E, \nep = 1 \)). The total bandwidth cost for \( i = 1 \) can be expressed as
\begin{equation*}
    \approx 2 \nparams (\ndp-1) + 2b \dmodel (\npp-1) + 8bL \sqrt{\dff \dmodel \ntp} = c_{\text{DP}} \ndp + c_{\text{PP}} \npp + c_{\text{TP}} \sqrt{\ntp} - d,
\end{equation*}
where \( c_{\text{DP}}, c_{\text{PP}}, c_{\text{TP}}, d \) are strictly positive constants depending on the model dimensions and batch size. In this problem, for \( \ngpu \) sufficiently large\footnote{When \( \ngpu \) is small, the constraint \( \ntp, \ndp, \npp \geq 1 \) will bind on the more expensive parallelism methods, so e.g. the optimal solution is likely to have \( \ntp = 1 \) over a wide range of \( \ngpu \) values. This is a big reason why this argument is too simplistic, but it still illustrates the general principle at work.}, it is optimal to scale all parallelism methods at the same time with the total cluster size \( \ngpu \), specifically as
\[ \ndp, \npp \propto \ngpu^{1/4}, \, \ntp \propto \ngpu^{1/2}. \]

This recalls our conclusion in Section \ref{sec:multidim-slicing} that slicing along multiple problem dimensions is more efficient than slicing along any single one.

In practice, network bandwidth is not the only constraint on distributed training: arithmetic intensity (i.e. memory bandwidth), communication latency, whether that communication can be overlapped with computation, bubble size management for pipeline parallelism, \textit{et cetera} are all relevant factors. These factors can alter the exact quantitative conclusions of the above argument, but the basic intuition that scaling all parallelism methods together is superior to scaling one in isolation generally remains true.

Table \ref{tab:parallelism-summary} also offers insight into some factors that limit the methods from scaling arbitrarily far. Each needs to slice either on one of the two model width dimensions \( \dmodel, \dff \) or the batch dimension \( b \). A guideline is that if the amount of computation needed to train a model scales faster than \( \dff \dmodel b \) (the inherently \textit{parallel} problem volume, in contrast to the inherently \textit{serial} dimensions of layer count and optimizer steps) as a model is scaled up, distributed training can hit a bottleneck. We make this guideline quantitative in Section \ref{sec:closed-form-biggest-training-run}.

\section{What constrains distributed training?}
\label{sec:constraints}

Having discussed the available parallelism methods, we now turn to the main subject of our paper: the obstacles we face if we try to scale some combination of these methods arbitrarily far. Because all these methods incur data movement costs, the arithmetic utilization of the GPUs can fall if a training run becomes bottlenecked by this data movement. We reiterate our key questions:

\questionslist

In this section, we give a separate account of each fundamental constraint that we identify and model.

\subsection{Data movement}

Distributed training can run up against data movement limits for two reasons: because we're moving too much data \textit{inside an individual GPU} or \textit{between different GPUs}. We consider each in turn.

\subsubsection{Intra-GPU data Movement}
\label{sec:intra-gpu-data-movement}

A typical GPU with tensor cores can compute much faster than it can move data to and from DRAM: for example, an NVIDIA H100 SXM \citep{dgx_h100_datasheet} can perform a theoretical maximum of \( 2 \times 10^{15} \flops \) ($1 \times 10^{15}$ multiply-accumulates (MAC)/s) at 8-bit precision during dense matrix multiplications, but has a DRAM memory bandwidth of only \( 3.35 \, \, \text{TB/s} \), for an \textbf{arithmetic intensity} (the ratio of the two) of $\approx 299$ MAC/byte. An 8-bit matrix multiplication of shape $(M, K) \times (K, N) \to (M \times N)$ must perform $MKN$ MAC and, under ideal caching, read the two input matrices and write the output matrix once each, for $MK + KN + MN$ bytes accessed total. For the H100's computational resources to be balanced, we thus must have $1/299 \approx (MK + KN + MN)/MNK = 1/M + 1/N + 1/K$. In the square case,
\begin{equation*}
    M = N = K \approx 896.
\end{equation*}
If the dimensions are substantially smaller, then the tensor cores must go underutilized as they will be bottlenecked on data movement to and from DRAM. Similar considerations apply at lower levels of the memory hierarchy as well. We discuss this further in Section \ref{sec:utilization-cliff} and Appendix \ref{sec:model-matmul}.

Distributed training across more GPUs or experts requires splitting the problem and hence reducing at least one of \( M, N, K \), worsening arithmetic intensity.

The main way to counter the degradation in arithmetic intensity is by increasing the dimensions of the matrix multiplications: at least one of \( \dmodel, \, \dff, \, b \) (Section \ref{sec:toymodel}), either making the model fatter and shorter, or increasing the batch size. Scaling \( b \) runs into the critical batch size limit, and it's not clear how far scaling model width \( \dmodel \) and \( \dff \) at the expense of model depth $L$ can go while remaining near the compute-optimal frontier. We discuss these matters in greater detail in Sections \ref{sec:critical-batch-size} and \ref{sec:layer-scaling}. The broad conclusion is there may be no free lunch for arithmetic intensity.

\subsubsection{Inter-GPU data movement}
\label{sec:network-bandwidth}

In addition to data movement inside individual GPUs, distributed training requires the movement of data between them. We quantified this cost in Section \ref{sec:parallelism}.

Each GPU has an upper bound to the rate of information it can receive or transmit per unit time. This means that asymptotically, the scaling of the total network bandwidth of a cluster can at most be linear in cluster size, even if the interconnect switches, wires, \textit{et cetera} are costless. In practice, scaling is often \textit{sublinear}. For instance, a high-bandwidth NVLink region may be limited to a single node: GPT-4 (\cite{openai2023gpt4}) used clusters of A100s \citep{patel2023gpt4} with an NVLink node size of 8. Increasing the tensor-parallel degree \( \ntp \) for such a cluster thus demands some fraction of bandwidth-hungry all-reduce communication over slower interconnects such as InfiniBand.

Even in an optimistic linear scaling regime where we do not have to rely in part on slower connections at other levels of the network hierarchy, scaling a cluster increases \textit{both} the arithmetic throughput \textit{and} the available network bandwidth proportionally, but the data movement \textit{per GPU} increases as we scale the degrees of parallelism, as seen in Section \ref{sec:parallelism}. Thus if we hold the model and batch size constant and simply scale up the training cluster, arithmetic intensity with regard to the network must eventually shrink to the point that this communication becomes a bottleneck.

Even if we scale these dimensions, the question remains whether they scale fast enough relative to total training compute of the model, as \( \ngpu \) must be at least proportional to the training run compute given constant utilization and training duration. The answer depends on factors discussed in Sections \ref{sec:critical-batch-size} and \ref{sec:layer-scaling}.

\subsection{The critical batch size}
\label{sec:critical-batch-size}

Because data and pipeline parallelism both involve slicing the data matrix along the batch dimension, increasing the batch size \( b \) helps with scaling up these two methods of parallelism without damaging arithmetic intensity. This is always doable, but only useful to the extent it reduces noise in the gradient estimate; at some scale, the estimate is precise enough that further noise reductions are not useful. \citep{shallue2019measuringeffectsdataparallelism}

In an ideal scaling regime, \( n \) gradient steps with a batch size \( b \) and learning rate \( \eta \) should reduce loss equivalently to taking a single gradient step with a batch size \( n \cdot b \) and learning rate \( n \cdot \eta \) \citep{mccandlish2018empirical}, so increasing the batch size effectively parallelizes the otherwise serial gradient descent steps. However, at some point, called the \textit{critical batch size}, this favorable scaling rapidly hits diminishing returns.

\cite{mccandlish2018empirical} conjecture, based on a second-order approximation, that a gradient step at batch size \( b \) should ideally reduce loss by
\begin{equation*}
    \Delta L = \frac{\Delta L_{\text{max}}}{1 + b_{\text{noise}}/b},
\end{equation*}
where \( \Delta L_{\text{max}} \) is the infinite batch size limit, and \( b_{\text{noise}} \) is a ``noise scale" at which the dependence of \( \Delta L \) on \( b \) goes from linear to sub-linear. In this model, the critical batch size \( b_{\text{crit}} \) is equal to the noise scale \( b_{\text{noise}} \).

They also conduct experiments and observe that models with smaller loss tend to have larger \( b_{\text{noise}} \) values, but holding loss constant, there is no impact of model size on \( b_{\text{noise}} \). As model loss decreases throughout training, this also means that \( b_{\text{noise}} \) increases throughout training, allowing larger batch sizes to be used later on in a training run compared to what is efficient at initialization.

The intuition for this is straightforward when we're using cross-entropy loss with respect to a ground truth distribution: a model with a smaller Kullback-Leibler divergence with the ground truth distribution can only be distinguished from ground truth by drawing more samples. Knowing which direction to go in for a useful gradient step certainly requires detecting that the model is not already optimal, so models with a smaller reducible loss per token can accommodate larger batch sizes (in units of tokens). This information-theoretic argument also makes the quantitative prediction that the noise scale \( b_{\text{noise}} \) should vary inversely proportionally with the reducible loss of the model, i.e. the Kullback-Leibler divergence of the model with the ground truth distribution.

If this conjecture is correct, and the dependence of the critical batch size on a dense model's properties factors through its reducible loss, then we can leverage the Chinchilla scaling law from \cite{hoffmann2022training} to predict how the critical batch size should scale with the training compute of a dense model. As the reducible loss of a Chinchilla-optimal dense model falls off like \( \propto 1/T^\alpha \) where \( T \) is the model's training compute and $\alpha \approx 1/6$, we would predict that the critical batch size should scale approximately with \( T^{1/6} \).

On the other hand, \cite{deepseekai2024deepseek} claim to find the scaling relationship \( b_{\text{crit}} \propto T^{0.33} \), although many of their models are overtrained relative to the Chinchilla law. For a fixed total training compute and batch size, an overtrained model must be \textit{smaller} (posing less opportunity for tensor parallelism at high arithmetic intensity) and \textit{train on more tokens}, hence use more serial steps, exacerbating latency bottlenecks (Section \ref{sec:latency}). If batch sizes can be scaled this aggressively even at the Chinchilla-optimal frontier, then opportunities for parallelism abound. Our baseline results employ the more conservative scaling assumption of \( b_{\text{crit}} \propto T^{1/6} \), though we analyze robustness to this assumption in Appendix \ref{sec:appendix-allowing-scaling-layer-batch}.

The above discussion relies on theoretical and empirical analysis of first-order optimization methods. While second-order methods have been investigated for years \citep{dauphin2014saddle,martens2015kfac,yao2020adahessian,liu2023sophia}, they pose significant scalability challenges: Full curvature information is normally intractable to calculate, so approximate and/or infrequent estimates must be used instead. Furthermore, the curvature (and its empirical estimate) can be very small or even negative and vary across the loss landscape, so that damping or trust-region approaches are required for stable learning. For these reasons, it remains unclear if second-order methods will ever be practical for frontier training runs. If they are, it's conceivable that critical batch sizes could greatly increase.

\subsection{Latency}
\label{sec:latency}

The training of a neural network involves some irreducible number of inherently serial operations, across model layers ($2L$ matrix multiplications in each direction in our toy model) and optimizer steps (with parallelism limits due to the critical batch size, as explained in Section \ref{sec:critical-batch-size}).

If lower bounds exist on the latency of communication between GPUs and matrix multiplication inside a GPU, then the sequential bottleneck in neural network training sets a lower bound for the duration of the entire training run, per Amdahl's law. In practice, bandwidth limits often become binding before latency limits, but latency limits are nevertheless significant because they are more difficult to overcome.

For example, one can always increase the total amount of bandwidth in a cluster by obtaining more GPUs and more network interconnects: in the bandwidth-bound regime this will reduce the arithmetic utilization of the entire cluster, but overall throughput will still increase. A typical result here might be a decay of \( \propto 1/\ngpu^{1/3} \) or \( \propto 1/\ngpu^{1/4} \) in the overall utilization rate, in a 3D- or 4D-parallel regime respectively (Section \ref{sec:parallelism}). In contrast, if the training process is latency-bound, further cluster scaling will not yield any additional benefits at all and the utilization rate will decay as \( \propto 1/\ngpu \). 

The two important sources of latency relevant during distributed training are:

\begin{enumerate}
    \item \textbf{Network latency:} Interconnects used in high-performance computing are optimized to have low latency, but even low-latency interconnects such as InfiniBand still require on the order of \( 1 \mus \) for a single point-to-point communication. Within a single node, the interconnect of choice for NVIDIA GPUs is NVLink, and a typical latency of an all-reduce operation that uses NVLink within a single node of 8 GPUs is on the order \( \approx 10 \mus \) (\cite{jeaugey2018nccl}).

    \item \textbf{Intra-GPU latency:} Even when working with a single GPU, there are sources of latency that constrain how quickly matrix multiplications may be performed on a single device. We say more about this in Section \ref{sec:absolute-limit}, but to summarize, we empirically measure a floor on matrix multiplication latency on an A100 using cuBLAS at around \( 4.5 \mus \).
\end{enumerate}

These latencies are much too small to matter for current training runs, as a typical duration for one matrix multiplication in a single GPU during the training of a current large language model is on the order of \( 1 \) to \( 10 \) milliseconds. For instance, Falcon-180B \citep{almazrouei2023falcon} used \( \dmodel = 12288 \), \( \dff = 4 \dmodel \),\( \ntpff = 8 \), \( b = 2^{22} \) and \( \ndp \cdot \npp = 512 \). If we assume e.g. 4 times as many microbatches as pipeline stages, their feedforward matrix multiplications on individual GPUs were of the shape \( 12288 \times 6144 \times 2048 \). On an A100, such a matrix multiplication takes \( \approx 1 \ms \) at full utilization.

However, as models become larger they need to be trained on more tokens \citep{hoffmann2022training} and the critical batch size might not increase proportionally (Section \ref{sec:critical-batch-size}). In this case, the time taken per gradient step has to fall in order that the entire training run still complete within an acceptable duration, and for very large models it could in principle fall below the timescales at which latency constraints would become binding. In Section \ref{sec:absolute-limit} we conclude that we have \( \approx 4 \oom \) more room to scale the training compute of state-of-the-art models before latencies begin to hurt utilization.

\subsection{Scaling the number of layers}
\label{sec:layer-scaling}

The need to increase the number of sequential operations in a neural network as the network is scaled up can have adverse effects on how parallelizable the training run of such a model can be, as model layers are inherently serialized. The effect of this is mitigated to some extent by techniques that reduce pipeline bubble sizes (Section \ref{sec:pipeline-parallelism}) and enable models to be effectively parallelized across the layer dimension. However, when network or GPU latency constraints become binding, avoiding additional serial steps in the forward and backward pass is critical for ensuring good utilization even when these techniques are used.

Unfortunately, there is little good research we've been able to find in the literature about what we call \textit{shape scaling laws}: how the performance of a model depends on the balance between its widths (as measured by \( \dmodel, \dff \)) and its depth (as measured by the number of layers \( L \)) \citep{kaplan2020scaling, alabdulmohsin2024gettingvitshapescaling}. The best we can do is back out implicit laws used in other work, and due to the clean nature of their data and the large number of data points we pick \cite{hoffmann2022training} as our reference in this section. Figure \ref{fig:layer-scaling-law} shows the results we obtain by analyzing the information about layer scaling in the models trained in \cite{hoffmann2022training}, with the best-fitting scaling law \( L \approx 3.67 \cdot (\nparams/10^6)^{0.27} \) where \( L \) is the number of layers and \( \nparams \) is the number of model parameters.

While this might be nothing more than a rule of thumb, we believe it is nevertheless an important finding because scaling a model by increasing the number of layers generally allows fewer opportunities for parallelism than increasing \( \dmodel \) or \( \dff \). This is because arithmetic intensity constrains the minimum dimensions of the individual matrix multiplications performed by a single device, and without scaling at least one of \( \dmodel, \dff\), or \( b \) the arithmetic intensity cannot effectively be improved. We will see the significance of this in Section \ref{sec:utilization-cliff}.

\section{Closed-form expressions for the biggest training run}
\label{sec:closed-form-biggest-training-run}

In this section and the next, we provide answers to the questions raised in Section \ref{sec:constraints}. We make some simplifications to derive analytic bounds here. After this preliminary analysis, we will introduce a complete model in Section \ref{sec:complete-model} which takes into account all of the constraints that we have previously discussed and closely matches our analytic results.

\subsection{The utilization cliff}
\label{sec:utilization-cliff}

Here we derive a closed-form expression upper-bounding the largest achievable training run, before data movement bottlenecks lower utilization substantially.

With even a modest degree of tensor parallelism, its bandwidth consumption greatly exceeds (Table \ref{tab:parallelism-summary}) the maximum possible point-to-point communication from pipeline or expert parallelism (e.g. the case $\npp \cdot i = L$), so we shall neglect their communication cost and take these at their maximum possible degrees $\npp = L$, $\nep = E$, restricting ourselves to considering only the all-reduce communication from data and tensor parallelism, across $\ndp \times \ntp$ groups (``workers") of $\npp \times \nep = LE$ GPUs each.

Recalling our discussion of multi-dimensional slicing in Section \ref{sec:multidim-slicing}, and taking $I = \dff, K = \dmodel, J = b/E$ in Eq. \ref{eq:allreduce-data-mvmt}, the all-reduce communication so imposed is equalized across all three dimensions and minimized when each worker has a cube-shaped unit of work $(d, d) \times (d, d) \to (d, d)$ for each matrix multiplication, with some side length
\begin{align}
    \label{eq:side-length}
    d = \dff/\ntpff = \dmodel/\ntpmodel = (b/E)/\ndp.
\end{align}
However, given our $\npp = L$ pipeline stages, we may assume at least $m \approx 2\npp = 2L$ microbatches, as this is the minimum to achieve the zero bubble condition under a ZB-H2 schedule (Section \ref{sec:pipeline-parallelism}). This means each actual ``physical" matrix multiplication onboard a single GPU corresponding to a single expert and pipeline stage uses a much smaller \textbf{nanobatch} size of\footnote{The nanobatch size $\bcrit = b/(E \cdot \ndp \cdot m)$ is the number of tokens seen at once by an individual GPU matrix multiplication kernel, taking into account routing into $E$ separate experts (whether or not on separate GPUs), data-parallel sharding of degree $\ndp$, and sharding into $m$ microbatches for pipeline paralellism.}
\begin{align}
    \label{eq:bprime}
    \bcrit = (b/E)/(\ndp \cdot m) = (b/E)/(\ndp \cdot 2L),
\end{align}
which is smaller than $d$ by the factor $2L$ if Eq. \ref{eq:side-length} holds. Under normal circumstances, this is already a problem for intra-GPU data movement to and from high-bandwidth memory (Section \ref{sec:intra-gpu-data-movement}), as this necessitates loading all of an expert's weights, and accumulating their gradients, once for each tiny nanobatch. However, as we are examining the limits to distributed training, we must consider the possibility that each of the $\ndp$ model replicas contains enough individual GPUs such that these weights, along with their gradients, can fit entirely in their aggregate SRAM, perhaps even in register banks. In such a case, their movement is required only once per batch to all-reduce gradients across model replicas, rather than once per nanobatch.

Even so, we cannot normally make the data-parallel degree $\ndp$ large enough to achieve the ideal of Eq. \ref{eq:side-length}, but instead just large enough that Eq. \ref{eq:bprime} equals some hardware-specific lower bound $\bcrit$, which we will estimate later using two separate methods. Then our data-parallel degree is
\begin{align}
\label{eq:ndp-bandwidth-constrained}
    \ndp = \frac b \bcrit \cdot \frac 1 {2 LE}.
\end{align}
Meanwhile, there is some minimum utilization-preserving weight submatrix size $\dcrit \times \dcrit$, which we will also estimate later. This makes our tensor-parallel degree $\ntp = \ntpff \times \ntpmodel$ equal to
\begin{equation}
    \label{eq:ntp-bandwidth-constrained}
    \ntp = \frac {\dff\dmodel} {\dcrit^2}.
\end{equation}
As discussed above, we assume the maximum possible pipeline- and expert-parallel degrees $\npp = L$, $\nep = E$, and take $\ndp$ and $\ntp$ as in Eqs. \ref{eq:ndp-bandwidth-constrained} and \ref{eq:ntp-bandwidth-constrained} so as to achieve the critical computation volume $\dcrit^2\bcrit$ per single-GPU matrix multiplication, with a total number of GPUs
\begin{align}
    \label{eq:ncrit}
    \ncrit &= \npp \cdot \nep \cdot \ndp \cdot \ntp = \frac 1 {4LE} \cdot \frac {b\nparams} {\dcrit^2 \bcrit},
\end{align}
where we have made use of the parameter count formula $\nparams = 2LE\dmodel\dff$ (Eq. \ref{eq:nparams}).

Assuming that a cluster of the critical size $\ncrit$ can run at close to full utilization by overlapping data movement with computation, its compute output over a time period \( \traintime \) is simply
\begin{align}
    \label{eq:training-compute-from-parallelism-and-time}
    \tcrit = \ncrit C \traintime,
\end{align}
where $C$ is the arithmetic throughput (MAC/second) of a single GPU at perfect utilization.

Bringing scaling laws into the argument, we now calculate the size of the model that can be trained in a given duration (e.g. 3 months) under the condition that utilization not seriously drop off. Chinchilla scaling laws \citep{hoffmann2022training} give us a compute-optimal training dataset size of roughly \( 20 \nparams \) tokens, with each training token requiring 3 MAC per parameter, so that the total compute required to train a model with \( \nparams \) parameters and sparsity factor $E$ is roughly
\begin{align}
    \label{eq:training-compute-from-params-and-sparsity}
    \tcrit = (60\mac) \cdot \nparams^2/E.
\end{align}
Setting Eqs. \ref{eq:training-compute-from-parallelism-and-time} and \ref{eq:training-compute-from-params-and-sparsity} equal and substituting in Eq. \ref{eq:ncrit} lets us solve for number of parameters:
\begin{align*}
    \nparams = \frac b L \cdot \frac{C \traintime}{(240\mac) \cdot {\dcrit^2 \bcrit}}.
\end{align*}
Taking this formula for $\nparams$ in Eq. \ref{eq:training-compute-from-params-and-sparsity},
\begin{align}
    \label{eq:ccrit}
    \tcrit &= \frac 1 {(960\mac) \cdot E} \left(\frac b L \cdot \frac{C \traintime}{{\dcrit^2 \bcrit}}\right)^2
\end{align}
is thus an upper bound on the critical training compute at which a model with sparsity factor $E$, a batch size of $b$, and a depth of $L$ MLP blocks can be trained over duration $\traintime$ on a cluster of GPUs whose specifications are held fixed, without loss of utilization. This formula is quadratically sensitive to most of its variables, and even quartically sensitive to the critical weight submatrix shape $\dcrit$, so any significant uncertainties can shift our estimate substantially. However, we can use it to place the utilization cliff within an order of magnitude or two.

\begin{wrapfigure}{R}{0.5\textwidth}
\vspace{-5pt}
\centering
  \includegraphics[width=0.48\textwidth]{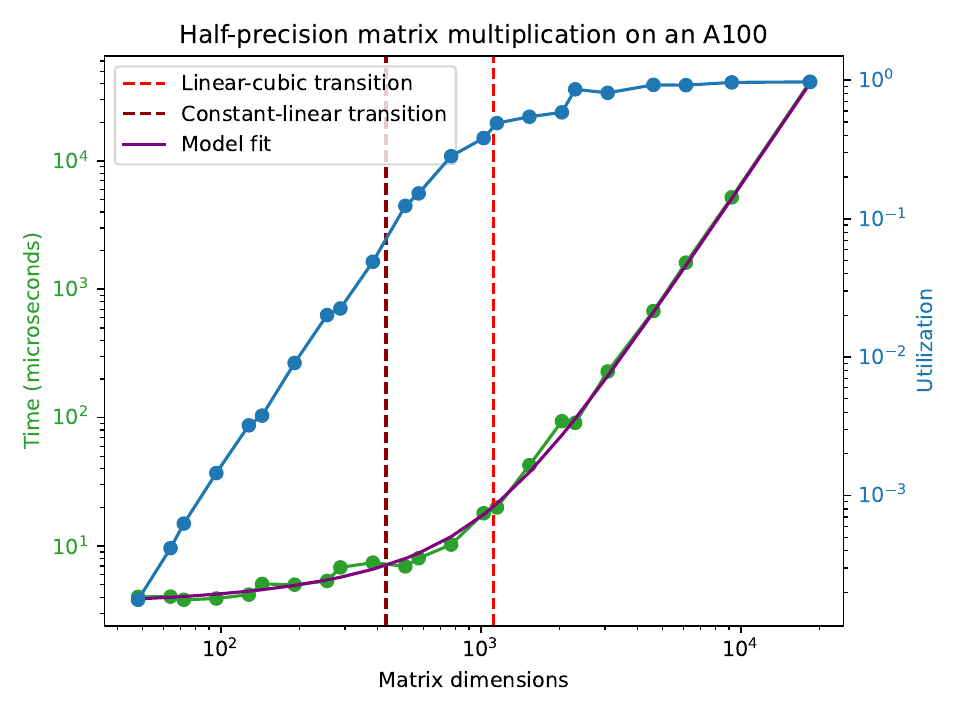}
  \caption{\small Default cuBLAS GEMM kernel performance on an A100 GPU. The green and blue curves show observed wall clock time and utilization, respectively. The purple curve shows a cubic polynomial fit to wall clock time, with dashed lines at the boundaries between a constant \( \approx 4.5 \mus \) latency-bound regime, a linear (likely occupancy-bound) regime, and a cubic compute-bound regime.}
  \label{fig:a100-cublas-plot}
\vspace{-30pt}
\end{wrapfigure}

In practice, we (weakly) expect the batch size $b$ and number of MLP blocks $L$ to scale similarly with training run size: specifically, batch size roughly as \( b \propto T^{1/6} \propto \nparams^{1/3} \) (Section \ref{sec:critical-batch-size}) and layer count roughly as \( L \propto \nparams^{0.27} \) (Section \ref{sec:layer-scaling}). Since Eq. \ref{eq:ccrit} depends on $b$ and $L$ only through their ratio, an approximation that they are constant may suffice. We will usually assume $b = 4 \times 10^6$ and $L = 100$, but explore variations on this assumption in Appendix \ref{sec:appendix-allowing-scaling-layer-batch}, concluding that under more optimistic, but speculative, assumptions, the critical compute threshold \( \tcrit \) might increase by about three orders of magnitude.

It is not clear how the sparsity factor \( E \) should be scaled. The most recent paper on scaling laws for mixture-of-experts (MoE) models known to us, \citep{krajewski2024scaling}, is agnostic on the question: the scaling law they derive does not incorporate \( E \) as a variable, instead using a fixed value \( E = 64 \) following recommendations from \cite{clark2022unifiedscalinglawsrouted}. It's also unknown to what degree doing so also permits further compute-efficient scaling of the batch size $b$. It is apparent from Eq. \ref{eq:ccrit} that if greater sparsity can enable larger batch sizes at a rate exceeding the square root of the sparsity factor, then sparsity permits larger training runs.

We'll now pursue two methods to estimate the critical nanobatch size $\bcrit$ and weight submatrix size $\dcrit \times \dcrit$, the first from latency bottlenecks, the second from bandwidth bottlenecks.

\subsubsection{From latencies}
\label{sec:cliff-from-latencies}

Definitionally, $\bcrit$ is the number of tokens in flight at any stage of its processing on a single GPU. So one lower bound for $\bcrit$ and $\dcrit$ comes from the fact that a GPU's unit of work $\dcrit^2\bcrit\mac$ for each of a nanobatch's matrix multiplications must take at least as long as the irreducible latencies from kernel launches and model- (i.e. tensor-, pipeline-, or expert-) parallel communication. If $t_L$ is the timescale of this latency, we therefore have:
\begin{align*}
    \frac {\dcrit^2\bcrit} C \ge \frac {t_L} \mac.
\end{align*}
Substituting into Eq. \ref{eq:ccrit},
\begin{align}
    \label{eq:ccrit-latency}
    T \le \frac {1 \mac} {960 \cdot E} \left(\frac b L \cdot \frac {\traintime} {t_L} \right)^2.
\end{align}
Figure \ref{fig:a100-cublas-plot} displays empirically observed matrix multiplication latencies on an A100, showing a \( \approx 4.5 \mus \) timescale below which the matrix multiplications cannot be accelerated due to various overheads of off-the-shelf cuBLAS kernels. Other GPUs are similar. Network latencies are also on the order of microseconds (Section \ref{sec:latency}) for an optimized all-reduce in which latency does not scale with the number of participants (e.g. a mesh network topology), so even a more optimized matrix multiplication kernel won't change our bottom line too much.\footnote{This does assume some network latency is incurred on a large fraction of MLP blocks, which is true so long as any tensor or expert parallelism is employed, or any significant degree of pipeline parallelism is employed. A purely data-parallel cluster would fail to fit model weights in DRAM, unless fully-sharded \citep{rajbhandari2020zero}, which would re-introduce network latencies at most layers.} To account for both factors, we take $t_L = 9 \mus$. Then given a training duration $\traintime$ of $3$ months, and typical values of $b = 4 \times 10^6, L = 100$, Eq. \ref{eq:ccrit-latency} for a dense model ($E = 1$) becomes
\begin{equation}
    \label{eq:ccrit-latency-concrete}
    \boxed{\mathbf{\tcrit = 3 \times 10^{30} \flop.}}
\end{equation}
We emphasize again that the quadratic sensitivity to uncertainty makes this an approximate estimate (within one or two orders of magnitude).

A discussion of possible methods to reduce $t_L$ is beyond our scope, but we note that a reduction by one order of magnitude, to $900 \ns$, would put it at only several times typical DRAM access latencies, despite the need for an all-reduce operation across inter-node interconnects, so this is likely a generous lower bound given anything like present interconnect technology. Given the quadratic role of $t_L$ in Eq. \ref{eq:ccrit-latency}, improved interconnects may therefore permit at most two orders of magnitude of further training compute.

\subsubsection{From bandwidth}
\label{sec:cliff-from-bandwidth}
Alternatively, we can consider utilization-preserving minimums imposed on $\bcrit$ and $\dcrit$ by network and memory bandwidth. Each GPU, for each nanobatch, performs two physical matrix multiplications on the forward pass of shape $(\dcrit, \dcrit) \times (\dcrit, \bcrit) \to (\dcrit, \bcrit)$, one for each layer of the expert and MLP block for which it's responsible. This entails $\dcrit^2\bcrit\mac$ twice in the forward pass and four times in the backward pass (Section \ref{sec:toymodel}), for $6\dcrit^2\bcrit\mac$ in total arithmetic. Furthermore, it participates in a tensor-parallel all-reduce across $\ntpff$ or $\ntpmodel$ peers of the $\dcrit \times \bcrit$ activation matrix (or its gradient) after each such multiplication, in both the forward and backward pass, receiving approximately $2\dcrit\bcrit$ words (Eq. \ref{eq:allreduce-words}) for each of four such all-reduces, or $8\dcrit\bcrit$ words total. This is balanced when
\begin{equation*}
    \frac{6\dcrit^2\bcrit\mac}{C} = \frac{8\dcrit\bcrit}{\Bnet},
\end{equation*}
where \( C \) is the GPU's arithmetic throughput in MAC per second, and \( \Bnet \) its unidirectional network bandwidth in words per second. This is solved by
\begin{equation}
\label{eq:dcrit-formula}
    \dcrit = \frac {4C} {3\Bnet} \cdot \frac 1 \mac.
\end{equation}
How large is the critical nanobatch size $\bcrit$? If the weight submatrices (and their gradients) fit in SRAM, the usual reason to require a large nanobatch, i.e. controlling DRAM bandwidth repeatedly accessing weights and their gradients (Section \ref{sec:intra-gpu-data-movement}), does not apply. This can potentially occur when the aggregate SRAM of a single model replica exceeds twice the model size:
\begin{align*}
    \text{weights in SRAM} \implies \frac \ngpu \ndp \cdot S \ge 2 N_p,
\end{align*}
where $S$ is the single-GPU SRAM word capacity. Using the substitution $\ngpu/\ndp = \npp \cdot \nep \cdot \ntp$, our choices above $\npp = L$, $\nep = E$, the tensor-parallel bandwidth bottleneck (Eq. \ref{eq:ntp-bandwidth-constrained}) $\ntp = \dff\dmodel/\dcrit^2$, and the parameter count formula (Eq. \ref{eq:nparams}) $\nparams = 2 LE \dmodel \dff$, we cancel $\nparams$ from both sides to get:
\begin{align*}
    \text{weights in SRAM} \implies \frac S {\dcrit^2} \ge 4.
\end{align*}
In this case, we set an aggressive lower bound of
\begin{align*}
    \bcrit = 16,\tag{if weights in SRAM}
\end{align*}
as tensor core instructions typically require each input dimension to be at least 8, and we assume at least twice as many tokens per nanobatch to support double-buffered data movement, ensuring that the tensor cores are well fed.

Otherwise, we make use of the fact that a typical weight gradient accumulation step entails $\bcrit \dcrit^2$ MAC and $\dcrit^2$ DRAM accesses in both directions: once to read the previously-accumulated gradient, and once to write the new gradient. Canceling $\dcrit^2$ from both sides, the DRAM bottleneck implies:
\begin{align*}
    \bcrit = \frac C \Bdram \cdot \frac 1 \mac,\tag{if weights in DRAM}
\end{align*}
where $\Bdram$ is the ``unidirectional" DRAM bandwidth (half the total bandwidth) in words per second.

Applying Eq. \ref{eq:ccrit} with these formulas, we now estimate the largest possible utilization-preserving training run for recent NVIDIA systems, considering only bandwidth bottlenecks. We do the analysis with a twist: at the \textit{node} level, treating the whole node as a single ``GPU". This allows us to identify the tightest constraints since the inter-node bandwidth bottleneck (InfiniBand except in the case of an H100 SuperPOD) is most severe. The results are in Table \ref{tab:nvidia-tcrit}. In most cases, the $\tcrit$ calculated by this method is two to three orders of magnitude below that given by the latency method, making bandwidth the operative constraint on utilization, and leaving only about two orders of magnitude of headroom above today's largest training runs \citep{EpochNotableModels2024}.
\begin{table}[b]
\centering
\begin{tabular}{l|l|l|l|l|l|l|l}
                      & \textbf{Arithmetic}         & \multicolumn{2}{l|}{\textbf{Unidir. bandwidth (words/s)}} & \textbf{SRAM}    &                   &                   & $\mathbf{\tcrit}$  \\ 
\textbf{GPU}          & \textbf{(MAC/s)}            & \textbf{Network}       & \textbf{DRAM}                 & \textbf{(words)} & $\dcrit$          & $\bcrit$          & \textbf{(FLOP)}    \\ \hline
DGX-1 (V100)          & $5.00 \times 10^{14}$       & \( 2.5 \times 10^{10} \)  & $1.8 \times 10^{12}$          & $151$M           & $26.7$k           & 278               & $1 \times 10^{27}$ \\
DGX A100              & $1.25 \times 10^{15}$       & \( 1.0 \times 10^{11} \)  & $3.1 \times 10^{12}$          & $366$M           & $16.7$k           & 401               & $3 \times 10^{28}$ \\
DGX H100              & $3.96 \times 10^{15}$       & \( 2.0 \times 10^{11} \)  & $6.7 \times 10^{12}$          & $487$M           & $26.4$k           & 591               & $2 \times 10^{28}$ \\
\ \ \ \ " in SuperPOD & "                           & \( 9.0 \times 10^{11} \)  & "                             & "                & $5.9$k            & 16                & $1 \times 10^{34}$ \\
\end{tabular}
\caption{FP16 specs of recent NVIDIA systems \citep{dgx1_v100_datasheet, dgx_a100_datasheet, dgx_h100_datasheet}, along with their corresponding critical weight matrix shape $\dcrit \times \dcrit$ as from Eq. \ref{eq:dcrit-formula} and critical nanobatch size $\bcrit$ (depending on whether weights and their gradients might fit in SRAM), and maximum dense ($E = 1$) three-month training run size $\tcrit$ as from Eq. \ref{eq:ccrit}, assuming $b = 4 \cdot 10^6, L = 100$ and considering only bandwidth bottlenecks.}
\label{tab:nvidia-tcrit}
\end{table}

The increased inter-node bandwidth in a SuperPOD allows for a factor of $\approx 20$ increase in tensor parallelism, which in turn allows model replicas to be large enough to fit weights and their gradients in SRAM, bypassing DRAM bandwidth bottlenecks and greatly reducing the nanobatch size. As such, the operative constraint in this case is latency (Eq. \ref{eq:ccrit-latency-concrete}).

\subsection{The absolute limit}
\label{sec:absolute-limit}

In the previous subsection, we pursued one approach to answering the first question from Section \ref{sec:constraints}, identifying the limits to scale imposed by latencies and bandwidth, assuming good utilization. Aggregate bandwidth and arithmetic throughput can in principle be scaled arbitrarily far, though perhaps with large decreases in utilization. But latency imposes hard limits to scale regardless of utilization. We consider those limits here.

If \( t_L \) is the latency of one matrix multiplication as in Section \ref{sec:cliff-from-latencies}, then a model forward and backward pass must take at least \( 4L t_L \) time no matter how many GPUs we have in the training cluster, for the two matrix multiplications required per each of $L$ MLP blocks in each direction. For a batch size \( b \) and Chinchilla-optimal dataset size \( D = 20\nparams \), this means the training duration must be at least \( 4L t_L \cdot (D/b) = 80 \nparams L t_L/b \). Setting this equal to the training timescale \( \traintime \) and solving for \( \nparams \),
\begin{align}
    \label{eq:ncrit-latency-bottlenecked}
\nparams = \frac b L \cdot \frac{\traintime}{80 t_L}
\end{align}
is the greatest number of parameters that can be trained over duration \( \traintime \).

Employing once again the Chinchilla equation \ref{eq:training-compute-from-params-and-sparsity} for the total training compute, and substituting Eq. \ref{eq:ncrit-latency-bottlenecked},
\begin{align}
    \label{eq:climit-latency-bottlenecked}
    T_\text{limit} = \frac {3\mac} {320 \cdot E} \left(\frac b L \cdot \frac{\traintime}{t_L}\right)^2
\end{align}
is the total compute for this latency-limited training run. Interestingly, this is exactly nine times as large as Eq. \ref{eq:ccrit-latency}, the latency-imposed limit on \textit{utilization-preserving} training runs. This factor of nine is attributable to the replacement of an \textit{efficient} pipeline schedule with a \textit{latency-minimizing} pipeline schedule.

Taking our usual values $b = 4 \times 10^6$, \( L = 100 \), \( \traintime = 3 \text{ months} \), \( t_L = 9 \mus \) gives
\begin{empheq}[box=\fbox]{align*}
    \mathbf{\nparams }&\mathbf{= 4 \times 10^{14},}\\
    \mathbf{\tlim }&\mathbf{= 2 \times 10^{31} \flop}
\end{empheq}
as the number of parameters and training compute of the largest dense model that can be trained over $\traintime$ duration, even at low utilization. We emphasize yet again that the quadratic sensitivity to uncertainty makes this an approximate estimate.

As in Section \ref{sec:utilization-cliff}, this estimate depends on $b$ and $L$ through their ratio. In Appendix \ref{sec:appendix-allowing-scaling-layer-batch}, we consider an optimistic scenario where the batch size can be scaled much faster than the layer count, in which \( \tlim \) could be as high as \( 3 \times 10^{36} \text{ FLOP} \), which is so large that a training run of that size would exceed annual world primary energy consumption. Whether latencies are of practical relevance or not, therefore, may heavily depend on how aggressively batch sizes can be scaled relative to the layer count of large models.

\section{Complete model and results}
\label{sec:complete-model}

\subsection{Model specification}
\label{sec:model-specification}

Our model has the following pieces:

\begin{enumerate}
    \item\label{foo} Assumptions about how the critical batch size \( b \), the training dataset size \( D \), and the model dimensions \( \dmodel, \dff, E, L \) should scale as models are increased in size.

    \item A model of matrix multiplications on a single GPU. This model incorporates HBM, L2, and L1 memory bandwidth bottlenecks, and the resulting under-utilization of streaming multiprocessors on NVIDIA GPUs when the matrix multiplications become unusually small.

    \item A hierarchical network description. We ignore network topology for the most part but take into account that different levels in a network hierarchy (e.g. intra-node and inter-node) will have different bandwidths and latencies at which they can facilitate communication across GPUs. For instance, such a description could look like ``we have an NVLink node size of 8, the NVLink bandwidth is \( 600 \gbs \) bidrectional per GPU with \( 10 \mus \) one-way latency, and inter-node communication takes place at \( 50 \gbs \) bidirectional per GPU with \( 5 \mus \) one-way latency."

    \item A method that calculates how long a training run of a specific model parallelized in a specific way across a cluster of \( \ngpu \) GPUs will take. Here, we take into account parameters such as the pipeline interleaving factor (Section \ref{sec:pipeline-interleaving}), whether zero-bubble pipeline parallelism is used (Section \ref{sec:zero-bubble-pp}), and how the 5D parallelism setup is divided across different levels of the network hierarchy.

    \item An optimizer that searches over all possible ways of parallelizing the training of a model across a fixed number \( \ngpu \) of GPUs and finds the one that takes the least time. If multiple strategies take minimal time, we break the symmetry by picking the one that has the least network communication time, even if this time can be hidden behind arithmetic.

    \item A search function that takes as input the training compute cost of a model, uses the scaling relations from (1) to infer its shape and critical batch size, looks for the smallest possible cluster size \( \ngpu \) satisfying certain divisibility constraints that (5) thinks can train this model in less than a given time \( \traintime \). Usually, we take \( \traintime \) to be 3 months.
\end{enumerate}

The full model therefore enables us to answer questions such as ``if we wished to train a \( 10^{27} \flop \) model in \( 4 \) months or less, what's the smallest number of GPUs we could accomplish this with, and what would be the parallelism setup minimizing training time and network communication costs?"

There are also effects that we omit:

\begin{itemize}
    \item GPU and node failures are ignored. We assume a cluster of idealized GPUs that never fail and that can always do computations and communication at constant rates. This biases our conclusions in an optimistic direction, as in practice the challenges of managing GPU failures in clusters at the million GPU scale or above are nontrivial but likely manageable.

    \item Activation recomputation is assumed unnecessary, as very large training runs typically require large clusters with more than enough memory to avoid it.
\end{itemize}

For a fuller description of our model, see Appendix \ref{sec:appendix-model-description}. The code we use to simulate it can also be found at \href{https://github.com/ege-erdil/limits_to_distributed_training}{this GitHub repository}.

\subsection{Results for current hardware and algorithms}
\label{sec:results-for-current-hardware}

We first simulate what we call our ``baseline scenario", considering DGX-1 (V100), DGX A100, and DGX H100 nodes under their datasheet configurations \citep{dgx1_v100_datasheet, dgx_a100_datasheet, dgx_h100_datasheet}, but with GPU clocks adjusted for expected thermal throttling based on empirical experiments with each GPU. Above the single-node NVLink network, there is a flat InfiniBand network.

The scaling relationships we use for these runs can be found in Appendix \ref{sec:model-scaling-assumptions}. The exact relations have a large influence on our results, as we've already seen in the closed-form results from Section \ref{sec:closed-form-biggest-training-run}.

\begin{figure}[h]
\centering
  \includegraphics[width=0.7\linewidth]{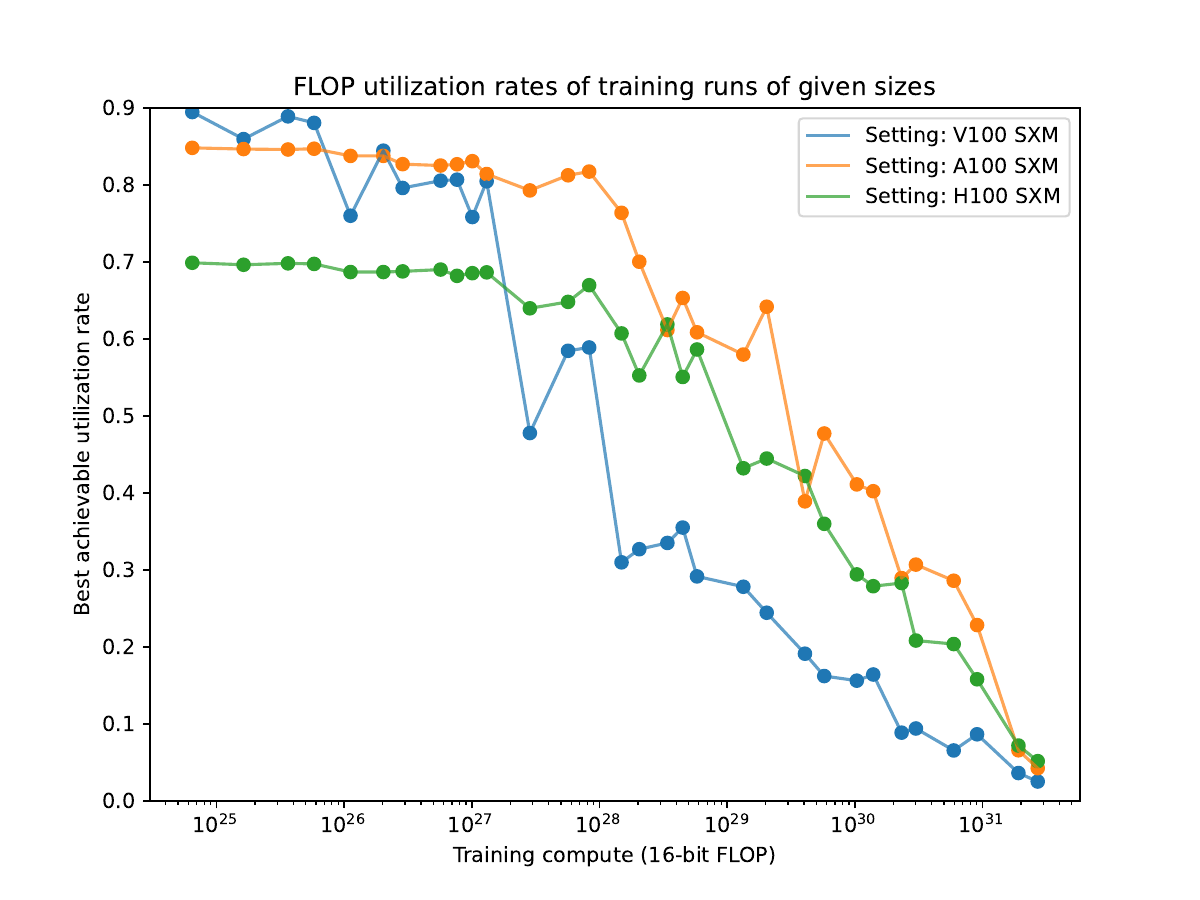}
  \caption{The MFU that's achieved by dense three-month training runs of different sizes when the cluster used to train them is made up of V100, A100, or H100 DGX nodes.}
  \label{fig:results-utilization-plot-1}
\end{figure}

\renewcommand{\arraystretch}{1.5}

\begin{table}[h]
\centering
\begin{tabular}{l|l|l}
                     & \textbf{For dense models}    & \textbf{For sparse models}    \\ \hline
V100 SXM             & \( 3 \times 10^{27} \) FLOP & \( 2 \times 10^{27} \) FLOP \\ \hline
A100 SXM             & \( 3 \times 10^{28} \) FLOP & \( 2 \times 10^{29} \) FLOP \\ \hline
H100 SXM             & \( 2 \times 10^{28} \) FLOP & \( 7 \times 10^{28} \) FLOP \\ \hline
\end{tabular}
\caption{A table summarizing when linear scaling of FLOP throughput with three month training run size stops in different cases, defined to be the point when model FLOP utilization (MFU; \cite{chowdhery2022palm}) falls below \( 80 \% \) of the utilization that a single GPU can achieve in sustained use.}
\label{tab:base-linear-scaling-summary}
\end{table}

Table \ref{tab:base-linear-scaling-summary} and Figure \ref{fig:results-utilization-plot-1} show the results of these simulations, which closely match our analytic results in Table \ref{tab:nvidia-tcrit}. Overall, \textbf{linear scaling breaks down around the $10^{27}$ to \( 10^{29} \flop \) scale}. In addition, as calculated in Section \ref{sec:absolute-limit}, \textbf{training runs past the \( 2 \times 10^{31} \flop \) scale are impossible due to latency constraints}, which is why the curves stop at that point. 

\begin{figure}[h]
\centering
  \includegraphics[width=1\linewidth]{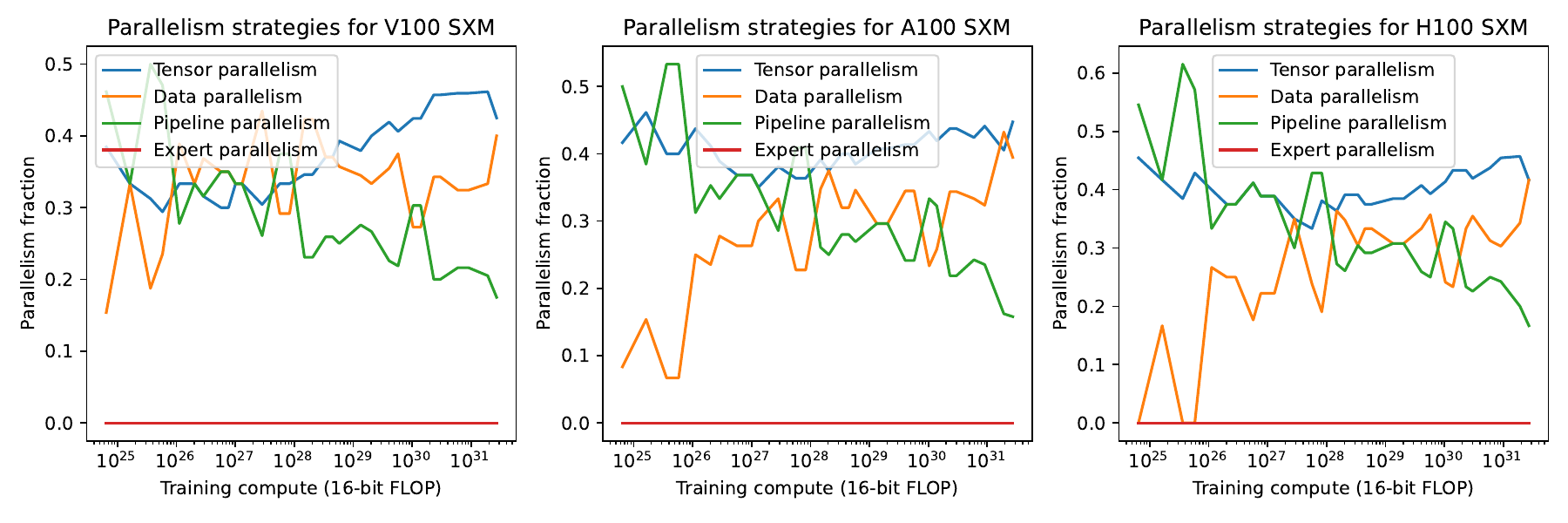}
  \caption{The fraction of overall parallelism that's dedicated to each parallelism strategy, for dense three-month training runs. The fraction for method \( X \) is computed as \( \log(N_X)/\log(\ngpu) \), i.e. the base \( \ngpu \) logarithm of \( N_X \). The fractions add up to \( 1 \) because the different parallelism degrees must multiply to the overall cluster size.}
  \label{fig:results-parallelism-plot-2}
\end{figure}

Figure \ref{fig:results-parallelism-plot-2} examines how training runs at different scales are parallelized. In general, as explained in Section \ref{sec:5d-parallelism}, it turns out to be optimal to scale all parallelism degrees together for large-scale training runs.

A few important phenomena that we've observed in simulations:

\begin{itemize}
    \item When memory bandwidth is the bottleneck, tensor parallelism is advantaged over data and pipeline parallelism due to its ability to slice along two dimensions at once.

    \item When network bandwidth is the bottleneck, pipeline parallelism is advantaged at small scales because it requires the least communication per parallelism degree. However, pipeline parallelism is costly due to pipeline bubbles unless a zero bubble scheme is used. At large scales, tensor parallelism is advantaged instead, because of the square root type scaling of communication costs derived in Section \ref{sec:tensor-parallelism}.
    
    \item When network latency is the bottleneck, data parallelism is advantaged over tensor and pipeline parallelism because it incurs the network latency cost twice per batch instead of multiple times per layer or once per pipeline stage. In addition, 2D tensor parallelism is often worse than 1D tensor parallelism because it incurs twice the network latency cost.
\end{itemize}

\subsection{Extending the linear scaling regime}

We've seen in Section \ref{sec:results-for-current-hardware} that using current hardware, it's not possible to extend the linear scaling regime past the \( 10^{29} \flop \) scale under our baseline scaling assumptions for dense models. This answers the first question we raised in the introduction. Now, we turn to answering the second question: what technological improvements are needed to extend the linear scaling regime past this scale?

There are two ways this can be achieved: by improving the hardware, or by improving the software (e.g. by finding ways to switch to more favorable scaling relations). We consider each in turn.

\subsubsection{Improving hardware}

Because the \( 10^{31} \flop \) scale is a limit set by latency constraints, going past this scale requires improving intra-GPU and network latency. Simply improving memory and network bandwidth is insufficient. However, there's still \( 3 \oom \) of room between \( 10^{28} \flop \) and \( 10^{31} \flop \), and the linear scaling regime can be extended in this limited window by increases in network or memory bandwidth, without any improvements to latency.

\begin{figure}[h]
\centering
  \includegraphics[width=0.75\linewidth]{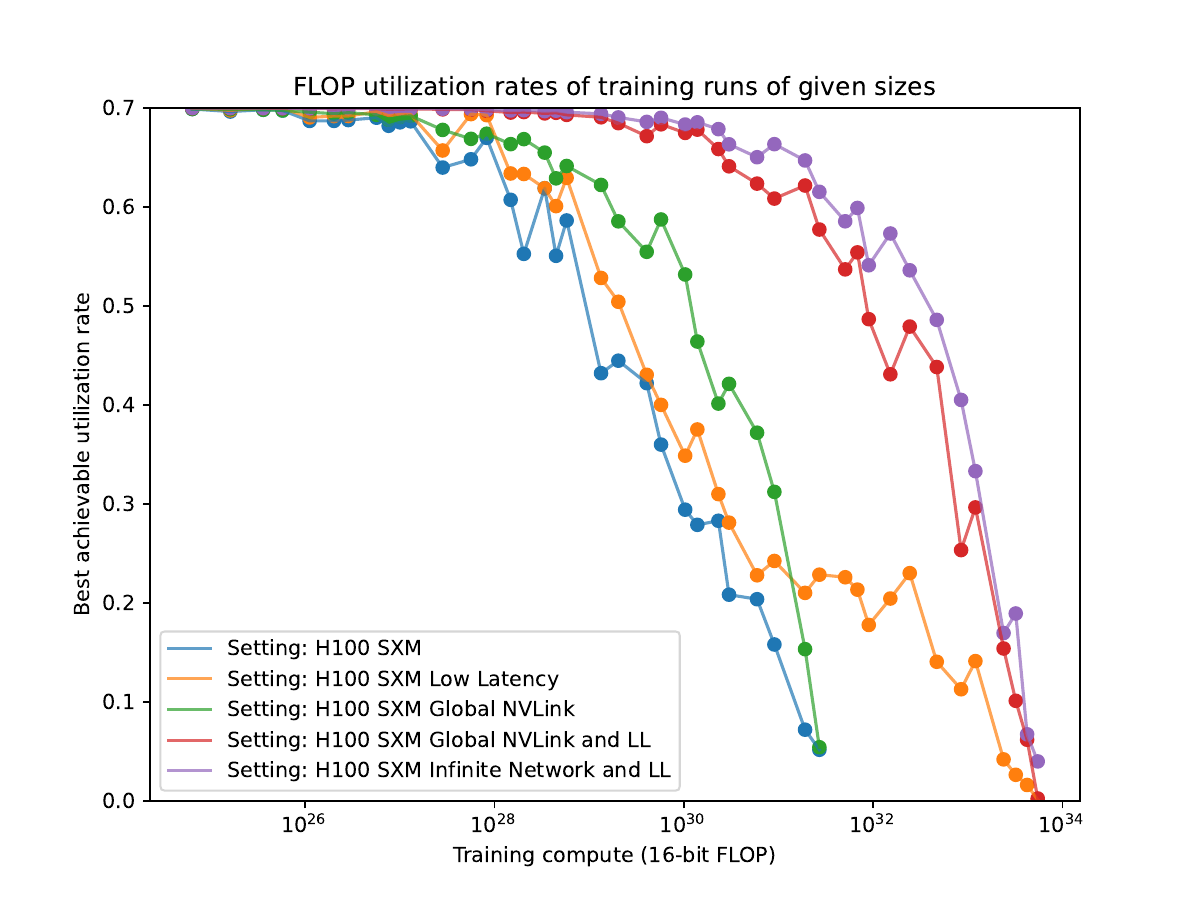}
  \caption{Utilization achieved by DGX H100s equipped with different hardware technologies.}
  \label{fig:linear-scaling-utilization-plot-1}
\end{figure}

Figure \ref{fig:linear-scaling-utilization-plot-1} shows the effect of relaxing different bottlenecks on utilization. Here, the 5 lines shown in the plot correspond to the following 5 sets of assumptions:

\begin{enumerate}
    \item \textbf{H100 SXM:} The baseline DGX H100 setup that consists of nodes of 8 GPUs connected over NVLink and different nodes connected over InfiniBand.
    \item \textbf{H100 SXM Low Latency:} Same as (1), except with all sources of latency (network and intra-GPU) divided by ten, which is perhaps the maximum reduction achievable without a major technological paradigm shift, as this puts inter-GPU latency at only several times the scale of a DRAM memory access.
    \item \textbf{H100 SXM Global NVLink:} A hypothetical network technology in which arbitrarily many H100s can be linked together at NVLink bandwidths.
    \item \textbf{H100 SXM Global NVLink and LL:} Same as (3), except with all sources of latency divided by ten.
    \item \textbf{H100 SXM Infinite Network and LL:} Same as (4), except network bandwidth is assumed infinite.
\end{enumerate}

The end of linear scaling for each case is listed in Table \ref{tab:linear-scaling-summary}, which also includes this information for sparse training runs. As usual, we emphasize that real-world results will differ somewhat as they are sensitive to exact achieved latencies and bandwidth, as well as assumptions about model depth and batch size scaling.

\renewcommand{\arraystretch}{1.5}

\begin{table}[h]
\centering
\begin{tabular}{l|l|l}
                     & \textbf{For dense models}    & \textbf{For sparse models}    \\ \hline
H100 SXM             & \( 2 \times 10^{28} \) FLOP & \( 7 \times 10^{28} \) FLOP \\ \hline
H100 SXM Low Latency    & \( 1 \times 10^{29} \) FLOP & \( 7 \times 10^{28} \) FLOP \\ \hline
H100 SXM Global NVLink    & \( 4 \times 10^{29} \) FLOP & \( 7 \times 10^{29} \) FLOP \\ \hline
H100 SXM Global NVLink and LL    & \( 5 \times 10^{31} \) FLOP & \( 1 \times 10^{32} \) FLOP \\ \hline
H100 SXM Infinite Network and LL    & \( 9 \times 10^{31} \) FLOP & \( 6 \times 10^{32} \) FLOP \\ \hline
\end{tabular}
\caption{The training run scale at which linear scaling of FLOP throughput with cluster size stops under different assumptions, defined to be the point at which the hardware utilization achieved in a training run falls below \( 80 \% \) of the utilization that a GPU can achieve in sustained use.}
\label{tab:linear-scaling-summary}
\end{table}

Figure \ref{fig:linear-scaling-utilization-plot-1} illustrates that a significant expansion of the linear scaling regime cannot be achieved solely by a reduction in latency or relative increase in network bandwidth, but requires their combination. With the (perhaps unrealistically) optimistic assumption of a $10\times$ reduction in latency and global NVLink-like bandwidth, the linear scaling regime is pushed out just over three orders of magnitude, to around $5 \times 10^{31} \flop$.

Even \( 10^{30} \flop \) is a very substantial compute budget: at September 2024 rental prices of around 3.5 USD per hour for the H100 SXM5 \citep{nebiusH100rental}, such a training run would cost over a trillion dollars. Our results can therefore also be interpreted in a more favorable light for the prospects of continued scaling: it may be possible to reach the \( 10^{30} \flop \) scale at good utilization, where economic rather than data movement considerations may dominate.

\subsubsection{Improving algorithms}

Algorithmic improvements offer another way to keep scaling past \( 10^{28} \flop \). In Section \ref{sec:closed-form-biggest-training-run}, we saw that the achievable training compute scales proportionally to \( (b/L)^2 \) where \( b \) is the global batch size (in tokens) and \( L \) is the number of MLP blocks in the model. While we assume a constant \( b/L \) ratio in that section, Appendix \ref{sec:appendix-allowing-scaling-layer-batch} considers the variable case and notes that changing the default batch size scaling law to the one from \cite{deepseekai2024deepseek} while holding the layer count scaling law fixed raises the achievable training compute by about three orders of magnitude.

\begin{figure}[h]
\centering
  \includegraphics[width=0.75\linewidth]{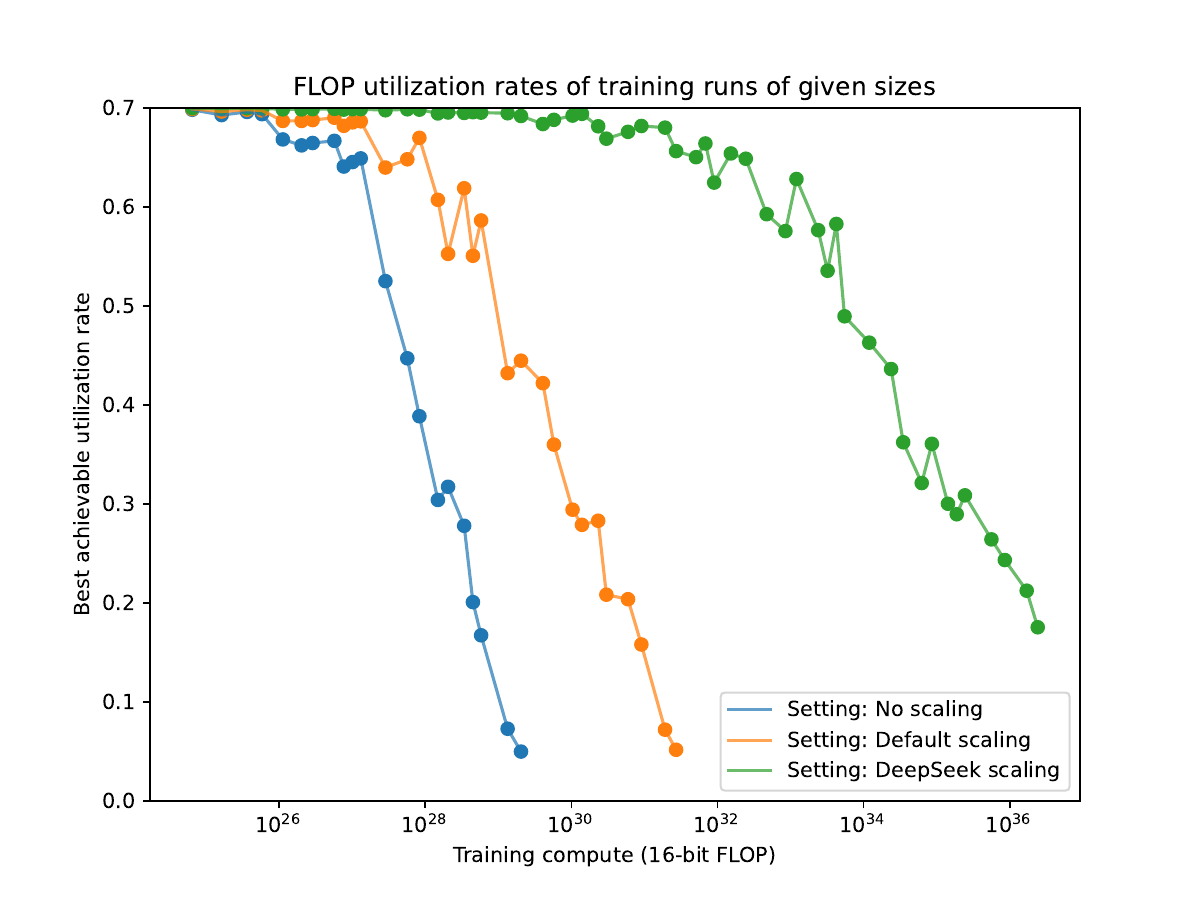}
  \caption{The H100 SXM utilization achieved by three-month dense training runs, under different critical batch size scaling laws.}
  \label{fig:linear-scaling-utilization-plot-2}
\end{figure}

This result is slightly more pronounced when we consider the full range of effects in our complete model, as shown in Figure \ref{fig:linear-scaling-utilization-plot-2}: quantitatively, the shift from default scaling to DeepSeek scaling raises the point at which linear scaling stops by \( 5.2 \oom \), from a training compute of \( 2 \times 10^{28} \flop \) to \( 3 \times 10^{33} \flop \). The ``no scaling" option in the plot corresponds to a fixed global batch size of \( 2^{22} \) tokens.

\section{Conclusion}
\label{sec:conclusion}

We now return to the two guiding questions we've set out to answer in this work. We repeat them once again for the sake of convenience:

\questionslist

Briefly summarizing our answers to these questions:

\begin{enumerate}
    \item Our best guess is \( \approx 2 \times 10^{28} \flop \). With aggressive batch size scaling \citep{deepseekai2024deepseek} this number could go up to \( \approx 3 \times 10^{33} \flop \), but we consider this to be rather optimistic. We think how quickly critical batch sizes can be scaled along with model sizes is a crucial question that has received little attention compared to its importance for continued scaling.
    
    \item In the future, improved interconnect bandwidth and latency might extend the linear scaling regime by at most two orders of magnitude. However, improvements in machine learning algorithms that allow for faster batch size scaling or shorter, fatter models have the potential to allow many more orders of magnitude of improvement.
\end{enumerate}

The recent turn of leading AI labs toward secrecy has made it difficult to acquire reliable information about algorithmic developments, which means much of the discussion of algorithmic issues in this paper has been speculative by necessity. It should be possible to substantially improve on our analysis with better information about:

\begin{enumerate}
    \item \textbf{Critical batch size scaling.} A better understanding of critical batch sizes, discussed in \cite{mccandlish2018empirical}, and how they scale with model size or reducible loss is essential. This is because making the batch size larger creates more opportunities for parallelism and reduces the number of gradient steps that must be taken in a single epoch during training. Compared to its importance, this question has received very little interest from researchers.

    \item \textbf{Model depth scaling and aspect ratio scaling laws.} Much of our work is based on the Chinchilla scaling law pioneered by \cite{hoffmann2022training}, which has proven very useful in understanding the limits to distributed training better. However, despite its utility, this scaling law does not make any predictions about how model depth ought to be scaled with model size and how much of a performance loss results from a sub-optimal choice. Because the model depth controls the number of sequential operations in a single forward pass, it has a big influence on how easy distributed training is at large scales. Consequently, we think the nature of the trade-off between model depth and model width deserves a more thorough investigation than we've been able to find in the literature.

    \item \textbf{Sparse model scaling laws.} Another shortcoming of the Chinchilla scaling law from \cite{hoffmann2022training} is that it only applies to dense models. With the growing popularity of mixture-of-experts models such as GPT-4 \citep{openai2023gpt4} and Mixtral \citep{jiang2024mixtral}, it has become important to understand the scaling laws of these models better as well. A useful start would be to check if the training compute-optimal dataset size of a sparse model scales approximately linearly with the total number of model parameters or not. There is also the possibility for interaction between sparsity and critical batch size.
\end{enumerate}

\pagebreak
\textbf{Acknowledgments}

We are grateful to Hailey Schoelkopf, Quentin Anthony, Daniel Ziegler, Deepak Narayanan, and Finbarr Timbers for their comments on an earlier draft of this paper; Robert Sandler for his assistance with illustrations; Jaime Sevilla and Tamay Besiroglu for comments on the paper and administrative assistance; Edu Roldan and Andrew Lucas for their work on the interactive tool; Maria de la Lama and Simon Jarvis for their assistance with publication; Mauricio Baker, Nicholas Brown, Felipe Calero Forero, Gabriel Kulp, Ashley Lin, and Aris Richardson for comments on a late draft; and Sihao Huang for his insights on hardware.

David Schneider-Joseph is very thankful to Lennart Heim for his mentorship and collaboration, as well as comments on this work; the Astra Fellowship under which some of this work took place; and the many great people at Constellation for thought-provoking discussions.

\bibliography{iclr2023_conference}

\begin{thebibliography}{48}
\providecommand{\natexlab}[1]{#1}
\providecommand{\url}[1]{\texttt{#1}}
\expandafter\ifx\csname urlstyle\endcsname\relax
  \providecommand{\doi}[1]{doi: #1}\else
  \providecommand{\doi}{doi: \begingroup \urlstyle{rm}\Url}\fi

\bibitem[Alabdulmohsin et~al.(2024)Alabdulmohsin, Zhai, Kolesnikov, and Beyer]{alabdulmohsin2024gettingvitshapescaling}
Ibrahim Alabdulmohsin, Xiaohua Zhai, Alexander Kolesnikov, and Lucas Beyer.
\newblock Getting {ViT} in shape: Scaling laws for compute-optimal model design, 2024.
\newblock URL \url{https://arxiv.org/abs/2305.13035}.

\bibitem[Almazrouei et~al.(2023)Almazrouei, Alobeidli, Alshamsi, Cappelli, Cojocaru, Debbah, Goffinet, Hesslow, Launay, Malartic, et~al.]{almazrouei2023falcon}
Ebtesam Almazrouei, Hamza Alobeidli, Abdulaziz Alshamsi, Alessandro Cappelli, Ruxandra Cojocaru, M{\'e}rouane Debbah, {\'E}tienne Goffinet, Daniel Hesslow, Julien Launay, Quentin Malartic, et~al.
\newblock The {Falcon} series of open language models.
\newblock \emph{arXiv preprint arXiv:2311.16867}, 2023.

\bibitem[Ba et~al.(2016)Ba, Kiros, and Hinton]{ba2016layernormalization}
Jimmy~Lei Ba, Jamie~Ryan Kiros, and Geoffrey~E. Hinton.
\newblock Layer normalization, 2016.
\newblock URL \url{https://arxiv.org/abs/1607.06450}.

\bibitem[Bi et~al.(2024)Bi, Chen, Chen, Chen, Dai, Deng, Ding, Dong, Du, Fu, Gao, Gao, Gao, Ge, Guan, Guo, Guo, Hao, Hao, He, Hu, Huang, Li, Li, Li, Li, Li, Liang, Lin, Liu, Liu, Liu, Liu, Liu, Liu, Lu, Lu, Luo, Ma, Nie, Pei, Piao, Qiu, Qu, Ren, Ren, Ruan, Sha, Shao, Song, Su, Sun, Sun, Tang, Wang, Wang, Wang, Wang, Wang, Wu, Wu, Xie, Xie, Xie, Xiong, Xu, Xu, Xu, Yang, You, Yu, Yu, Zhang, Zhang, Zhang, Zhang, Zhang, Zhang, Zhang, Zhang, Zhao, Zhao, Zhou, Zhou, Zhu, and Zou]{deepseekai2024deepseek}
Xiao Bi, Deli Chen, Guanting Chen, Shanhuang Chen, Damai Dai, Chengqi Deng, Honghui Ding, Kai Dong, Qiushi Du, Zhe Fu, Huazuo Gao, Kaige Gao, Wenjun Gao, Ruiqi Ge, Kang Guan, Daya Guo, Jianzhong Guo, Guangbo Hao, Zhewen Hao, Ying He, Wenjie Hu, Panpan Huang, Erhang Li, Guowei Li, Jiashi Li, Yao Li, Y.~K. Li, Wenfeng Liang, Fangyun Lin, A.~X. Liu, Bo~Liu, Wen Liu, Xiaodong Liu, Xin Liu, Yiyuan Liu, Haoyu Lu, Shanghao Lu, Fuli Luo, Shirong Ma, Xiaotao Nie, Tian Pei, Yishi Piao, Junjie Qiu, Hui Qu, Tongzheng Ren, Zehui Ren, Chong Ruan, Zhangli Sha, Zhihong Shao, Junxiao Song, Xuecheng Su, Jingxiang Sun, Yaofeng Sun, Minghui Tang, Bingxuan Wang, Peiyi Wang, Shiyu Wang, Yaohui Wang, Yongji Wang, Tong Wu, Y.~Wu, Xin Xie, Zhenda Xie, Ziwei Xie, Yiliang Xiong, Hanwei Xu, R.~X. Xu, Yanhong Xu, Dejian Yang, Yuxiang You, Shuiping Yu, Xingkai Yu, B.~Zhang, Haowei Zhang, Lecong Zhang, Liyue Zhang, Mingchuan Zhang, Minghua Zhang, Wentao Zhang, Yichao Zhang, Chenggang Zhao, Yao Zhao, Shangyan Zhou, Shunfeng Zhou, Qihao Zhu,
  and Yuheng Zou.
\newblock {DeepSeek LLM}: Scaling open-source language models with longtermism, 2024.

\bibitem[Chowdhery et~al.(2022)Chowdhery, Narang, Devlin, Bosma, Mishra, Roberts, Barham, Chung, Sutton, Gehrmann, Schuh, Shi, Tsvyashchenko, Maynez, Rao, Barnes, Tay, Shazeer, Prabhakaran, Reif, Du, Hutchinson, Pope, Bradbury, Austin, Isard, Gur-Ari, Yin, Duke, Levskaya, Ghemawat, Dev, Michalewski, Garcia, Misra, Robinson, Fedus, Zhou, Ippolito, Luan, Lim, Zoph, Spiridonov, Sepassi, Dohan, Agrawal, Omernick, Dai, Pillai, Pellat, Lewkowycz, Moreira, Child, Polozov, Lee, Zhou, Wang, Saeta, Diaz, Firat, Catasta, Wei, Meier-Hellstern, Eck, Dean, Petrov, and Fiedel]{chowdhery2022palm}
Aakanksha Chowdhery, Sharan Narang, Jacob Devlin, Maarten Bosma, Gaurav Mishra, Adam Roberts, Paul Barham, Hyung~Won Chung, Charles Sutton, Sebastian Gehrmann, Parker Schuh, Kensen Shi, Sasha Tsvyashchenko, Joshua Maynez, Abhishek Rao, Parker Barnes, Yi~Tay, Noam Shazeer, Vinodkumar Prabhakaran, Emily Reif, Nan Du, Ben Hutchinson, Reiner Pope, James Bradbury, Jacob Austin, Michael Isard, Guy Gur-Ari, Pengcheng Yin, Toju Duke, Anselm Levskaya, Sanjay Ghemawat, Sunipa Dev, Henryk Michalewski, Xavier Garcia, Vedant Misra, Kevin Robinson, Liam Fedus, Denny Zhou, Daphne Ippolito, David Luan, Hyeontaek Lim, Barret Zoph, Alexander Spiridonov, Ryan Sepassi, David Dohan, Shivani Agrawal, Mark Omernick, Andrew~M. Dai, Thanumalayan~Sankaranarayana Pillai, Marie Pellat, Aitor Lewkowycz, Erica Moreira, Rewon Child, Oleksandr Polozov, Katherine Lee, Zongwei Zhou, Xuezhi Wang, Brennan Saeta, Mark Diaz, Orhan Firat, Michele Catasta, Jason Wei, Kathy Meier-Hellstern, Douglas Eck, Jeff Dean, Slav Petrov, and Noah Fiedel.
\newblock {PaLM}: Scaling language modeling with pathways, 2022.

\bibitem[Clark et~al.(2022)Clark, de~las Casas, Guy, Mensch, Paganini, Hoffmann, Damoc, Hechtman, Cai, Borgeaud, van~den Driessche, Rutherford, Hennigan, Johnson, Millican, Cassirer, Jones, Buchatskaya, Budden, Sifre, Osindero, Vinyals, Rae, Elsen, Kavukcuoglu, and Simonyan]{clark2022unifiedscalinglawsrouted}
Aidan Clark, Diego de~las Casas, Aurelia Guy, Arthur Mensch, Michela Paganini, Jordan Hoffmann, Bogdan Damoc, Blake Hechtman, Trevor Cai, Sebastian Borgeaud, George van~den Driessche, Eliza Rutherford, Tom Hennigan, Matthew Johnson, Katie Millican, Albin Cassirer, Chris Jones, Elena Buchatskaya, David Budden, Laurent Sifre, Simon Osindero, Oriol Vinyals, Jack Rae, Erich Elsen, Koray Kavukcuoglu, and Karen Simonyan.
\newblock Unified scaling laws for routed language models, 2022.
\newblock URL \url{https://arxiv.org/abs/2202.01169}.

\bibitem[Dauphin et~al.(2014)Dauphin, Pascanu, G{\"{u}}l{\c{c}}ehre, Cho, Ganguli, and Bengio]{dauphin2014saddle}
Yann~N. Dauphin, Razvan Pascanu, {\c{C}}aglar G{\"{u}}l{\c{c}}ehre, Kyunghyun Cho, Surya Ganguli, and Yoshua Bengio.
\newblock Identifying and attacking the saddle point problem in high-dimensional non-convex optimization.
\newblock \emph{CoRR}, abs/1406.2572, 2014.
\newblock URL \url{http://arxiv.org/abs/1406.2572}.

\bibitem[Devlin et~al.(2019)Devlin, Chang, Lee, and Toutanova]{devlin2019bertpretrainingdeepbidirectional}
Jacob Devlin, Ming-Wei Chang, Kenton Lee, and Kristina Toutanova.
\newblock {BERT}: Pre-training of deep bidirectional transformers for language understanding, 2019.
\newblock URL \url{https://arxiv.org/abs/1810.04805}.

\bibitem[Dubey et~al.(2024)Dubey, Jauhri, Pandey, Kadian, Al-Dahle, Letman, Mathur, Schelten, Yang, Fan, et~al.]{dubey2024llama3herdmodels}
Abhimanyu Dubey, Abhinav Jauhri, Abhinav Pandey, Abhishek Kadian, Ahmad Al-Dahle, Aiesha Letman, Akhil Mathur, Alan Schelten, Amy Yang, Angela Fan, et~al.
\newblock The {Llama} 3 herd of models, 2024.
\newblock URL \url{https://arxiv.org/abs/2407.21783}.

\bibitem[{Epoch AI}(2024)]{EpochNotableModels2024}
{Epoch AI}.
\newblock Data on notable {AI} models, 2024.
\newblock URL \url{https://epochai.org/data/notable-ai-models}.
\newblock Accessed: 2024-10-09.

\bibitem[Fedus et~al.(2022)Fedus, Zoph, and Shazeer]{fedus2022switchtransformersscalingtrillion}
William Fedus, Barret Zoph, and Noam Shazeer.
\newblock Switch transformers: Scaling to trillion parameter models with simple and efficient sparsity, 2022.
\newblock URL \url{https://arxiv.org/abs/2101.03961}.

\bibitem[Graham et~al.(2016)Graham, Bureddy, Lui, Rosenstock, Shainer, Bloch, Goldenerg, Dubman, Kotchubievsky, Koushnir, et~al.]{graham2016scalable}
Richard~L Graham, Devendar Bureddy, Pak Lui, Hal Rosenstock, Gilad Shainer, Gil Bloch, Dror Goldenerg, Mike Dubman, Sasha Kotchubievsky, Vladimir Koushnir, et~al.
\newblock Scalable hierarchical aggregation protocol ({SHArP}): A hardware architecture for efficient data reduction.
\newblock In \emph{2016 First International Workshop on Communication Optimizations in HPC (COMHPC)}, pp.\  1--10. IEEE, 2016.

\bibitem[Gu \& Dao(2023)Gu and Dao]{gu2023mamba}
Albert Gu and Tri Dao.
\newblock Mamba: Linear-time sequence modeling with selective state spaces.
\newblock \emph{arXiv preprint arXiv:2312.00752}, 2023.

\bibitem[He et~al.(2015)He, Zhang, Ren, and Sun]{he2015deepresiduallearningimage}
Kaiming He, Xiangyu Zhang, Shaoqing Ren, and Jian Sun.
\newblock Deep residual learning for image recognition, 2015.
\newblock URL \url{https://arxiv.org/abs/1512.03385}.

\bibitem[Hestness et~al.(2017)Hestness, Narang, Ardalani, Diamos, Jun, Kianinejad, Patwary, Yang, and Zhou]{hestness2017scaling}
Joel Hestness, Sharan Narang, Newsha Ardalani, Gregory~F. Diamos, Heewoo Jun, Hassan Kianinejad, Md. Mostofa~Ali Patwary, Yang Yang, and Yanqi Zhou.
\newblock Deep learning scaling is predictable, empirically.
\newblock \emph{CoRR}, abs/1712.00409, 2017.
\newblock URL \url{http://arxiv.org/abs/1712.00409}.

\bibitem[Hoffmann et~al.(2022)Hoffmann, Borgeaud, Mensch, Buchatskaya, Cai, Rutherford, Casas, Hendricks, Welbl, Clark, et~al.]{hoffmann2022training}
Jordan Hoffmann, Sebastian Borgeaud, Arthur Mensch, Elena Buchatskaya, Trevor Cai, Eliza Rutherford, Diego de~Las Casas, Lisa~Anne Hendricks, Johannes Welbl, Aidan Clark, et~al.
\newblock Training compute-optimal large language models.
\newblock \emph{arXiv preprint arXiv:2203.15556}, 2022.

\bibitem[Huang et~al.(2019)Huang, Cheng, Bapna, Firat, Chen, Chen, Lee, Ngiam, Le, Wu, and Chen]{huang2019gpipe}
Yanping Huang, Youlong Cheng, Ankur Bapna, Orhan Firat, Dehao Chen, Mia Chen, HyoukJoong Lee, Jiquan Ngiam, Quoc~V Le, Yonghui Wu, and zhifeng Chen.
\newblock {GPipe}: Efficient training of giant neural networks using pipeline parallelism.
\newblock In H.~Wallach, H.~Larochelle, A.~Beygelzimer, F.~d\textquotesingle Alch\'{e}-Buc, E.~Fox, and R.~Garnett (eds.), \emph{Advances in Neural Information Processing Systems}, volume~32. Curran Associates, Inc., 2019.
\newblock URL \url{https://proceedings.neurips.cc/paper_files/paper/2019/file/093f65e080a295f8076b1c5722a46aa2-Paper.pdf}.

\bibitem[Jeauguey(2018)]{jeaugey2018nccl}
Sylvain Jeauguey.
\newblock Multi-{GPU} training with {NCCL}.
\newblock \textsc{url:}~\url{https://on-demand.gputechconf.com/gtc/2018/presentation/s8462-multi-gpu-training-with-nccl.pdf}, 2018.
\newblock Presented at NVIDIA GTC 2018.

\bibitem[Jiang et~al.(2024{\natexlab{a}})Jiang, Sablayrolles, Roux, Mensch, Savary, Bamford, Chaplot, de~las Casas, Hanna, Bressand, Lengyel, Bour, Lample, Lavaud, Saulnier, Lachaux, Stock, Subramanian, Yang, Antoniak, Scao, Gervet, Lavril, Wang, Lacroix, and Sayed]{jiang2024mixtral}
Albert~Q. Jiang, Alexandre Sablayrolles, Antoine Roux, Arthur Mensch, Blanche Savary, Chris Bamford, Devendra~Singh Chaplot, Diego de~las Casas, Emma~Bou Hanna, Florian Bressand, Gianna Lengyel, Guillaume Bour, Guillaume Lample, Lélio~Renard Lavaud, Lucile Saulnier, Marie-Anne Lachaux, Pierre Stock, Sandeep Subramanian, Sophia Yang, Szymon Antoniak, Teven~Le Scao, Théophile Gervet, Thibaut Lavril, Thomas Wang, Timothée Lacroix, and William~El Sayed.
\newblock Mixtral of experts, 2024{\natexlab{a}}.

\bibitem[Jiang et~al.(2024{\natexlab{b}})Jiang, Lin, Zhong, Huang, Chen, Zhang, Peng, Li, Xie, Nong, Jia, He, Chen, Bai, Hou, Yan, Zhou, Sheng, Jiang, Xu, Wei, Zhang, Nie, Zou, Zhao, Xiang, Liu, Li, Jia, Ye, Jin, and Liu]{jiang2024megascale}
Ziheng Jiang, Haibin Lin, Yinmin Zhong, Qi~Huang, Yangrui Chen, Zhi Zhang, Yanghua Peng, Xiang Li, Cong Xie, Shibiao Nong, Yulu Jia, Sun He, Hongmin Chen, Zhihao Bai, Qi~Hou, Shipeng Yan, Ding Zhou, Yiyao Sheng, Zhuo Jiang, Haohan Xu, Haoran Wei, Zhang Zhang, Pengfei Nie, Leqi Zou, Sida Zhao, Liang Xiang, Zherui Liu, Zhe Li, Xiaoying Jia, Jianxi Ye, Xin Jin, and Xin Liu.
\newblock {MegaScale}: Scaling large language model training to more than 10,000 {GPUs}, 2024{\natexlab{b}}.

\bibitem[Kaplan et~al.(2020)Kaplan, McCandlish, Henighan, Brown, Chess, Child, Gray, Radford, Wu, and Amodei]{kaplan2020scaling}
Jared Kaplan, Sam McCandlish, Tom Henighan, Tom~B Brown, Benjamin Chess, Rewon Child, Scott Gray, Alec Radford, Jeffrey Wu, and Dario Amodei.
\newblock Scaling laws for neural language models.
\newblock \emph{arXiv preprint arXiv:2001.08361}, 2020.

\bibitem[Krajewski et~al.(2024)Krajewski, Ludziejewski, Adamczewski, Pi{\'o}ro, Krutul, Antoniak, Ciebiera, Kr{\'o}l, Odrzyg{\'o}{\'z}d{\'z}, Sankowski, et~al.]{krajewski2024scaling}
Jakub Krajewski, Jan Ludziejewski, Kamil Adamczewski, Maciej Pi{\'o}ro, Micha{\l} Krutul, Szymon Antoniak, Kamil Ciebiera, Krystian Kr{\'o}l, Tomasz Odrzyg{\'o}{\'z}d{\'z}, Piotr Sankowski, et~al.
\newblock Scaling laws for fine-grained mixture of experts.
\newblock \emph{arXiv preprint arXiv:2402.07871}, 2024.

\bibitem[Kumar et~al.(1994)Kumar, Grama, Gupta, and Karypis]{kumar1994introduction}
Vipin Kumar, Ananth Grama, Anshul Gupta, and George Karypis.
\newblock \emph{Introduction to parallel computing}, volume 110.
\newblock Benjamin/Cummings Redwood City, CA, 1994.

\bibitem[Lepikhin et~al.(2020)Lepikhin, Lee, Xu, Chen, Firat, Huang, Krikun, Shazeer, and Chen]{lepikhin2020gshard}
Dmitry Lepikhin, HyoukJoong Lee, Yuanzhong Xu, Dehao Chen, Orhan Firat, Yanping Huang, Maxim Krikun, Noam Shazeer, and Zhifeng Chen.
\newblock {GShard}: Scaling giant models with conditional computation and automatic sharding.
\newblock \emph{arXiv preprint arXiv:2006.16668}, 2020.

\bibitem[Lewis et~al.(2021)Lewis, Bhosale, Dettmers, Goyal, and Zettlemoyer]{lewis2021baselayers}
Mike Lewis, Shruti Bhosale, Tim Dettmers, Naman Goyal, and Luke Zettlemoyer.
\newblock {BASE} layers: Simplifying training of large, sparse models.
\newblock \emph{CoRR}, abs/2103.16716, 2021.
\newblock URL \url{https://arxiv.org/abs/2103.16716}.

\bibitem[Liu et~al.(2023)Liu, Li, Hall, Liang, and Ma]{liu2023sophia}
Hong Liu, Zhiyuan Li, David Hall, Percy Liang, and Tengyu Ma.
\newblock Sophia: {A} scalable stochastic second-order optimizer for language model pre-training.
\newblock \emph{CoRR}, abs/2305.14342, 2023.
\newblock \doi{10.48550/ARXIV.2305.14342}.
\newblock URL \url{https://doi.org/10.48550/arXiv.2305.14342}.

\bibitem[Martens \& Grosse(2015)Martens and Grosse]{martens2015kfac}
James Martens and Roger~B. Grosse.
\newblock Optimizing neural networks with {Kronecker}-factored approximate curvature.
\newblock \emph{CoRR}, abs/1503.05671, 2015.
\newblock URL \url{http://arxiv.org/abs/1503.05671}.

\bibitem[McCandlish et~al.(2018)McCandlish, Kaplan, Amodei, and Team]{mccandlish2018empirical}
Sam McCandlish, Jared Kaplan, Dario Amodei, and OpenAI~Dota Team.
\newblock An empirical model of large-batch training, 2018.

\bibitem[Narayanan et~al.(2021)Narayanan, Shoeybi, Casper, LeGresley, Patwary, Korthikanti, Vainbrand, Kashinkunti, Bernauer, Catanzaro, et~al.]{narayanan2021efficient}
Deepak Narayanan, Mohammad Shoeybi, Jared Casper, Patrick LeGresley, Mostofa Patwary, Vijay Korthikanti, Dmitri Vainbrand, Prethvi Kashinkunti, Julie Bernauer, Bryan Catanzaro, et~al.
\newblock Efficient large-scale language model training on {GPU} clusters using {Megatron-LM}.
\newblock In \emph{Proceedings of the International Conference for High Performance Computing, Networking, Storage and Analysis}, pp.\  1--15, 2021.

\bibitem[Nebius(2024)]{nebiusH100rental}
Nebius.
\newblock {GPU} prices: {Nvidia H100, A100}, 2024.
\newblock URL \url{https://web.archive.org/web/20240915193540/https://nebius.ai/prices}.
\newblock Accessed: 21 October 2024.

\bibitem[NVIDIA(2019)]{dgx1_v100_datasheet}
NVIDIA.
\newblock Nvidia {DGX-1}: The essential instrument of {AI} research.
\newblock \url{https://www.nvidia.com/content/dam/en-zz/Solutions/Data-Center/dgx-1/dgx-1-rhel-datasheet-nvidia-us-808336-r3-web.pdf}, 2019.
\newblock Accessed: 2024-10-18.

\bibitem[NVIDIA(2020)]{dgx_a100_datasheet}
NVIDIA.
\newblock {NVIDIA DGX A100}, 2020.
\newblock URL \url{https://resources.nvidia.com/en-us-dgx-systems/dgx-ai}.

\bibitem[NVIDIA(2022)]{dgx_h100_datasheet}
NVIDIA.
\newblock {NVIDIA DGX H100}, 2022.
\newblock URL \url{https://resources.nvidia.com/en-us-dgx-systems/ai-enterprise-dgx}.

\bibitem[{NVIDIA}(2022)]{nvidia_hopper_2022}
{NVIDIA}.
\newblock {NVIDIA H100} tensor core {GPU} architecture.
\newblock \url{https://resources.nvidia.com/en-us-tensor-core/gtc22-whitepaper-hopper}, 2022.
\newblock Accessed: 2024-10-09.

\bibitem[NVIDIA(2024)]{nvidia_cutlass_documentation}
NVIDIA.
\newblock {CUTLASS} documentation.
\newblock \url{https://github.com/NVIDIA/cutlass/wiki/Documentation}, 2024.
\newblock Accessed: 2024-10-09.

\bibitem[OpenAI(2023)]{openai2023gpt4}
OpenAI.
\newblock {GPT-4} technical report, 2023.

\bibitem[Patel \& Wong(2023)Patel and Wong]{patel2023gpt4}
Dylan Patel and Gerald Wong.
\newblock {GPT-4} architecture, infrastructure, training dataset, costs, vision, {MoE}, 7 2023.
\newblock URL \url{https://www.semianalysis.com/p/gpt-4-architecture-infrastructure?utm_campaign=post&utm_medium=web}.

\bibitem[Peng et~al.(2023)Peng, Alcaide, Anthony, Albalak, Arcadinho, Cao, Cheng, Chung, Grella, GV, et~al.]{peng2023rwkv}
Bo~Peng, Eric Alcaide, Quentin Anthony, Alon Albalak, Samuel Arcadinho, Huanqi Cao, Xin Cheng, Michael Chung, Matteo Grella, Kranthi~Kiran GV, et~al.
\newblock {RWKV}: Reinventing {RNNs} for the transformer era.
\newblock \emph{arXiv preprint arXiv:2305.13048}, 2023.

\bibitem[Qi et~al.(2024)Qi, Wan, Huang, and Lin]{qi2023zero}
Penghui Qi, Xinyi Wan, Guangxing Huang, and Min Lin.
\newblock Zero bubble pipeline parallelism.
\newblock \emph{CoRR}, abs/2401.10241, 2024.
\newblock \doi{10.48550/ARXIV.2401.10241}.
\newblock URL \url{https://doi.org/10.48550/arXiv.2401.10241}.

\bibitem[Rajbhandari et~al.(2020)Rajbhandari, Rasley, Ruwase, and He]{rajbhandari2020zero}
Samyam Rajbhandari, Jeff Rasley, Olatunji Ruwase, and Yuxiong He.
\newblock {ZeRO}: Memory optimizations toward training trillion parameter models.
\newblock In \emph{{SC20}: International Conference for High Performance Computing, Networking, Storage and Analysis}, pp.\  1--16. IEEE, 2020.

\bibitem[Sevilla et~al.(2022)Sevilla, Heim, Ho, Besiroglu, Hobbhahn, and Villalobos]{sevilla2022compute}
Jaime Sevilla, Lennart Heim, Anson Ho, Tamay Besiroglu, Marius Hobbhahn, and Pablo Villalobos.
\newblock Compute trends across three eras of machine learning, 2022.

\bibitem[Shallue et~al.(2019)Shallue, Lee, Antognini, Sohl-Dickstein, Frostig, and Dahl]{shallue2019measuringeffectsdataparallelism}
Christopher~J. Shallue, Jaehoon Lee, Joseph Antognini, Jascha Sohl-Dickstein, Roy Frostig, and George~E. Dahl.
\newblock Measuring the effects of data parallelism on neural network training, 2019.
\newblock URL \url{https://arxiv.org/abs/1811.03600}.

\bibitem[Shazeer et~al.(2017)Shazeer, Mirhoseini, Maziarz, Davis, Le, Hinton, and Dean]{shazeer2017outrageouslylargeneuralnetworks}
Noam Shazeer, Azalia Mirhoseini, Krzysztof Maziarz, Andy Davis, Quoc Le, Geoffrey Hinton, and Jeff Dean.
\newblock Outrageously large neural networks: The sparsely-gated mixture-of-experts layer, 2017.
\newblock URL \url{https://arxiv.org/abs/1701.06538}.

\bibitem[Vaswani et~al.(2017)Vaswani, Shazeer, Parmar, Uszkoreit, Jones, Gomez, Kaiser, and Polosukhin]{vaswani2017attention}
Ashish Vaswani, Noam Shazeer, Niki Parmar, Jakob Uszkoreit, Llion Jones, Aidan~N Gomez, {\L}ukasz Kaiser, and Illia Polosukhin.
\newblock Attention is all you need.
\newblock In I.~Guyon, U.~Von Luxburg, S.~Bengio, H.~Wallach, R.~Fergus, S.~Vishwanathan, and R.~Garnett (eds.), \emph{Advances in Neural Information Processing Systems}, volume~30. Curran Associates, Inc., 2017.
\newblock URL \url{https://proceedings.neurips.cc/paper_files/paper/2017/file/3f5ee243547dee91fbd053c1c4a845aa-Paper.pdf}.

\bibitem[Wang \& Komatsuzaki(2021)Wang and Komatsuzaki]{wang2021gpt}
Ben Wang and Aran Komatsuzaki.
\newblock {GPT-J-6B}: A 6 billion parameter autoregressive language model, 2021.

\bibitem[Weng \& Brockman(2022)Weng and Brockman]{openai2022largemodels}
Lilian Weng and Greg Brockman.
\newblock Techniques for training large neural networks, 2022.
\newblock URL \url{https://openai.com/research/techniques-for-training-large-neural-networks}.
\newblock Accessed: 2023-12-01.

\bibitem[Yao et~al.(2020)Yao, Gholami, Shen, Keutzer, and Mahoney]{yao2020adahessian}
Zhewei Yao, Amir Gholami, Sheng Shen, Kurt Keutzer, and Michael~W. Mahoney.
\newblock {ADAHESSIAN:} an adaptive second order optimizer for machine learning.
\newblock \emph{CoRR}, abs/2006.00719, 2020.
\newblock URL \url{https://arxiv.org/abs/2006.00719}.

\bibitem[Zhou et~al.(2022)Zhou, Lei, Liu, Du, Huang, Zhao, Dai, Chen, Le, and Laudon]{zhou2022expertchoice}
Yanqi Zhou, Tao Lei, Hanxiao Liu, Nan Du, Yanping Huang, Vincent~Y. Zhao, Andrew~M. Dai, Zhifeng Chen, Quoc Le, and James Laudon.
\newblock Mixture-of-experts with expert choice routing.
\newblock \emph{CoRR}, abs/2202.09368, 2022.
\newblock URL \url{https://arxiv.org/abs/2202.09368}.

\end{thebibliography}
\bibliographystyle{iclr2023_conference}

\section{Appendices}

\subsection{Appendix: Detailed model description}
\label{sec:appendix-model-description}

\subsubsection{Matrix multiplication on a single device}
\label{sec:model-matmul}

A naive model of matrix multiplication on one device is to assume perfect utilization, in which case the time taken for a matrix multiplication of dimensions \( m \times k \times n \) is \( mkn/C \) where \( C \) is the MAC per second that the device can perform. Our model improves on this in three ways:

\begin{enumerate}
    \item We consider the memory bandwidth required for the matrix multiplication at different levels of the memory hierarchy. A big matrix needs to be read from HBM to L2, from L2 to shared memory (physically the same SRAM as L1 cache), and from shared memory into the registers. If communication is perfectly overlapped with computation inside a single device, the time taken for the matrix multiplication will be the maximum of the time taken for the arithmetic and the time taken for the data movement at all levels.

    \item There is a minimum latency to each matrix multiplication due to setup and teardown of the CUDA kernel along with memory access latency.

    \item Due to thermal throttling, devices often fall short of the peak boosted clock speeds reported in official data sheets. We correct this by performing empirical benchmarks on various GPUs to observe what clock speeds devices can sustain during realistic workloads.
\end{enumerate}

The complicated part of the model is (1). To understand how we model this, consider the memory hierarchy levels involved in performing a matrix multiplication on an NVIDIA GPU using a typical kernel design, in which responsibility for calculating the output matrix is successively tiled into smaller sub-rectangles at the streaming multiprocessor (SM) and processing block (executing one or more ``warps" of 32 threads each) level. Because of the correspondingly smaller output matrix size at each smaller level, the IO intensity of the input reads is increased. In detail \citep{nvidia_cutlass_documentation}:
\begin{enumerate}
    \item The input matrices must be read from HBM into L2.
    \item he necessary pieces of the input matrices for an SM's output tile are loaded from L2 (or distributed shared memory in later GPU generations \citep{nvidia_hopper_2022} if another SM has already loaded the relevant data) into that SM's shared memory. This and other data movement typically happens asynchronously and double-buffered so as to overlap with computation.
    \item Each of the four processing blocks on an SM performs coalesced reads of the input data from the SM's shared memory into its threads' registers.
    \item Small chunks of input data move from the processing blocks' register banks to the tensor core, where the arithmetic of a warp-level matrix multiply-accumulate operation is performed, and the result accumulated in registers in that processing block which have been pre-allocated for the corresponding tile of the output matrix.
    \item After many repetitions of the above steps, all arithmetic has been performed and the output tiles fully accumulated. The result is finally written back through the memory hierarchy and into HBM.
\end{enumerate}

In the above sequence, each step has an input-output intensity which depends on the extent to which the tiled matrix has to be tiled at that level in the hierarchy. Other sequences are possible, based on different tiling strategies. We always assume tiling by the weight (or weight gradient) matrix dimensions, even if this is not the output matrix, as this usually allows for the smallest memory intensity. We calculate this intensity at the L2, SM, and warp level of the hierarchy, which tells us how much data movement is required at each level for the amount of computation involved in the matrix multiplication. We divide these by associated memory bandwidths obtained from official documentation or micro-benchmarks to estimate how much time is required for communication at each level. Taking a maximum of these values and the arithmetic time, then adding kernel setup/teardown latency, tells us how long the matrix multiplication takes.

\subsubsection{Network communication}
\label{sec:model-network-communication}

A straightforward communication model for a network with a single level of hierarchy proceeds by taking the quantitative estimates from Section \ref{sec:parallelism}. For a training step taken on a single batch, we can estimate the following network communication costs:

\begin{itemize}
    \item \textbf{Data parallelism:} \( \nparams \) words all-reduced across \( \ndp \) ranks per batch.
    \item \textbf{Tensor parallelism:} \( \dff \) words all-reduced across \( \ntpmodel \) ranks and \( \dmodel \) words all-reduced across \( \ntpff \) ranks twice per layer per token, once each for the forward and the backward pass.
    \item \textbf{Pipeline parallelism:} \( \dmodel \) words communicated point-to-point \( 2(\npp \cdot i - 1) \) times per token, where \( i \) is the pipeline interleaving factor, and the factor of two counts both the forward and the backward pass.
    \item \textbf{Expert parallelism:} \( \dmodel \) additional words communicated point-to-point an average of \( 2 \cdot (1 - 1/\nep) \cdot (L - \npp \cdot i) \) times per token. The factor of \( 1 - 1/\nep \) is the probability that a request must be routed to a different expert than the current one, while \( (L - \npp \cdot i) \) is the number of layer boundaries that are not pipeline communication boundaries.
\end{itemize}

Letting the bidirectional data movement costs per batch be denoted by \( M_{\text{DP}}, M_{\text{TP}}, M_{\text{PP}}, M_{\text{EP}} \) respectively, and assuming that we have a per-GPU bidirectional network bandwidth of \( \bnetwork \), we can calculate the total time necessary for inter-GPU data movement as
\[ t_{\text{network}} = t_{\text{network, DP}} + t_{\text{network, TP}} + t_{\text{network, PP}} + t_{\text{network, EP}} = \frac{M_{\text{DP}} + M_{\text{TP}} + M_{\text{PP}} + M_{\text{EP}}}{\ngpu \cdot \bnetwork} \]
We assume that this communication, except for that imposed by data parallelism or during pipeline startup and clearing, can overlap with GPU arithmetic.

When there are several different levels of network hierarchy (e.g. nodes of \( 8 \) GPUs with \( \bnetwork = 900 \gbs \) per GPU, superpods of \( 32 \) nodes with \( \bnetwork = 450 \gbs \) per GPU and InfiniBand across superpods achieving \( \bnetwork = 100 \gbs \) per GPU), matters become more complicated:

\begin{enumerate}
    \item We must decide how different degrees of parallelism are partitioned across different levels in the network hierarchy. For example, because tensor parallelism is more communication-intensive than data and pipeline parallelism, it usually makes sense to do tensor parallelism at lower levels of the network hierarchy (e.g. inside nodes using NVLink for internal communication) and data, pipeline, or expert parallelism at higher levels.

    \item All-reduces and point-to-point communications will have to be implemented hierarchically to make efficient use of the available network bandwidth.

    \item Pipeline parallelism and expert parallelism interact in a nontrivial way when multiple levels of network hierarchy are taken into account. Specifically, having to send requests to an expert outside of a lower level of the network hierarchy can cause information to be routed across slower connections even if pipeline parallelism is present at lower levels of the network hierarchy. This is taken into account in our model.
\end{enumerate}

We address each of these points in order. First, if our network has \( H \) levels of hierarchy enumerated as \( h \in \{ 1, 2, \ldots, H \} \), and each level \( h \) of the hierarchy has a per-GPU bandwidth of \( \bnetwork(h) \) where levels with smaller \( h \) have faster bandwidth, we assume there's some partition of the parallelism degrees across the \( H \) levels. Specifically, if \( X \) is a dimension of parallelism, we assume there is a factorization \( N_X = N_X(1) \cdot N_X(2)\cdot \ldots \cdot N_X(H) \).

In such a setup, we can perform a hierarchical all-reduce of \( d \) words across dimension \( X \) as follows:

\begin{itemize}
    \item Reduce \( d \) words across \( N_X(1) \) participants \( \prod_{h=2}^H N_X(h) \) times using a per-GPU bandwidth of \( \bnetwork(1) \).
    \item Reduce \( d \) words across \( N_X(2) \) participants \( \prod_{h=3}^H N_X(h) \) times using a per-GPU bandwidth of \( \bnetwork(2) \).
    \item \ldots
    \item Reduce \( d \) words across \( N_X(H) \) participants once using a per-GPU bandwidth of \( \bnetwork(H) \).
\end{itemize}

This explains how to modify the calculation for data and tensor parallelism: each all-reduce operation is replaced by \( H \) all-reduce operations in ascending order across the network hierarchy, with bandwidth, though not the latency, of these all-reduces overlapped.

For pipeline and expert parallelism, we must think more carefully because of the interaction between these two modes of parallelism: a single point-to-point communication can simultaneously do the job for both. We consider each combination \( (h_{\text{PP}}, h_{\text{EP}}) \in \{0, \ldots, H\}^2 \) of network hierarchy levels possibly requiring communication from pipeline and expert parallelism (with level $0$ corresponding to the case where no communication is required), determine the frequency of that combination, and then tally a communication on the highest level of the two $h' = \max\{h_{\text{PP}}, h_{\text{EP}}\}$ with that frequency.

\textbf{Pipeline-parallel communication frequencies.} Define $\ppp(h)$ to be the probability that an inter-layer interface (i.e., between two MLP blocks) requires pipeline-parallel communication at level $h$ of the network hierarchy.

If $\npp = 1$ (i.e. no pipeline parallelism is utilized), then clearly $\ppp(0) = 1$ and $\ppp(h) = 0$ for $h > 0$, since $h = 0$ corresponds to the case that no pipeline-parallel communication is required.

If $\npp > 1$ (i.e. any pipeline parallelism at all is utilized), let $h^* = \max\{h : \npp(h) > 1\}$ be the highest level of the network hierarchy at which pipeline parallelism ever takes place. For a given level $h$, there are
\begin{equation}
    \label{eq:pipeline-stage-interfaces-at-or-above}
    \left(i \prod_{k=h}^{h^*} \npp(k)\right) - 1
\end{equation}
pipeline stage interfaces at level $h$ or above, where $i$ is the interleaving factor as in Section \ref{sec:pipeline-interleaving}.

When $h = h^*$, this is just $i \cdot \npp(h^*) - 1$, and hence the frequency that one of the $L - 1$ inter-layer interfaces is an $h^*$-level pipeline stage interface is just
\begin{equation*}
    \ppp(h^*) = \frac {i \cdot \npp(h^*) - 1}{L - 1}.
\end{equation*}
Otherwise when $1 \le h < h^*$, we employ Eq. \ref{eq:pipeline-stage-interfaces-at-or-above} twice to see that there are
\begin{equation*}
    \left(i \prod_{k=h}^{h^*} \npp(k)\right) - \left(i \prod_{k=h+1}^{h^*} \npp(k)\right) = \left(i \prod_{k=h+1}^{h^*} \npp(k)\right) \left( \npp(h) - 1 \right)
\end{equation*}
pipeline stage interfaces at exactly level $h$, yielding the frequency
\begin{equation*}
\ppp(h) = \left(i \prod_{k=h+1}^{h^*} \npp(k)\right) \cdot \frac  {\npp(h) - 1} {L - 1}\tag{$1 \le h < h^*$}
\end{equation*}
that an inter-layer interface requires pipeline-parallel communication at this level of the network hierarchy.

Employing Eq. \ref{eq:pipeline-stage-interfaces-at-or-above} one last time with $h = 1$ yields the total number $i\npp - 1$ of pipeline stage interfaces across all levels of the network hierarchy. Subtracting from $L - 1$ gives us the number of inter-layer interfaces that do \textit{not} require pipeline communication, with frequency
\begin{equation*}
    \ppp(0) = \frac {L - i \npp} {L - 1}.
\end{equation*}
\textbf{Expert-parallel communication frequencies.} Now similarly define $\pep(h)$ to be the probability that one of the $L - 1$ inter-layer interfaces requires expert-parallel communication at level $h$ of the network hierarchy.

We assume expert routing is uniform, so the probability is
\begin{equation}
    \label{eq:expert-same-rank-prob}
    \left(\prod_{k=h}^H N_{\text{EP}}(k)\right)^{-1}
\end{equation}
that the current and next expert lie at the same rank on all levels of the network hierarchy at or above $h$. Employing Eq. \ref{eq:expert-same-rank-prob} twice, the probability is
\begin{equation}
    \label{eq:expert-routing-prob}
    \pep(h) = \left(\prod_{k=h+1}^H N_{\text{EP}}(k)\right)^{\mathclap{-1}} - \left(\prod_{k=h}^H N_{\text{EP}}(k)\right)^{\mathclap{-1}} = \frac {N_{\text{EP}}(h) - 1} {\prod_{k=h}^H N_{\text{EP}}(k)}\tag{$h \ge 1$}
\end{equation}
that the current and next expert lie at the same rank on all levels of the network hierarchy above $h$ but not at level $h$, and hence that expert-parallel communication must occur on exactly level $h$. As a sanity check, observe that the sum
\[ \sum_{h=1}^H \pep(h) \]
telescopes to $1 - 1 / \nep$, where the leftover probability of \( 1/\nep \) corresponds to the chance that no expert-parallel communication will be needed, as the next expert is stored in the same rank as the current expert on all levels of the network hierarchy. Thus \( \pep(0) = 1/\nep \).

\textbf{Joint communication frequencies.} Activations (and their gradients, on the backward pass) must be communicated at the inter-layer interface between MLP blocks on the level of the network hierarchy corresponding to the maximum of that required for pipeline parallelism and that required for expert parallelism. Since we assume expert routing is independent of layer and token, the probability that communication is required on network level $h'$ is:
\begin{equation*}
    \pptop(h') = \sum_{\mathclap{\substack{0 \le h_\text{PP}, h_\text{EP} \le H \\ \max\{h_\text{PP}, h_\text{EP}\} = h'}}} \ppp(h_\text{PP})\pep(h_\text{EP}).
\end{equation*}
\textbf{Latencies and overlap.} Two final wrinkles: we will need to make assumptions about to what extent network communication is overlapped with computation, and also account for network latencies. To this end, we decompose the network communication time as
\[ t_{\text{network}} = t_{\text{overlapped network}} + t_{\text{nonoverlapped network}} \]
and assume the first term can be overlapped with computations while the second term cannot be. In the ideal case we have \( t_{\text{nonoverlapped network}} = 0 \), but we also find it useful to consider cases where overlapping is not quite perfect.

On top of the above calculations regarding network bandwidth, we must also track network latency. We do this by considering the number of times that each all-reduce or point-to-point communication operation must occur serially (as opposed to in parallel) and multiplying this number by a network latency parameter at that particular level of the network hierarchy. \( t_{\text{network latency}} \) is then defined as the sum of all of these latency timescales. Furthermore, for the latency (as opposed to bandwidth) of peer-to-peer communication, we always assume the worst-case for expert routing (i.e. $\pep(h^*) = 1$ where $h^*$ is the highest level of the network hierarchy employing expert parallelism, and $\pep(h) = 0$ for $h \neq h^*$), as all tokens in a microbatch must wait on the slowest one.

\subsubsection{Pipeline bubbles}
\label{sec:model-bubbles}

We take pipeline bubbles into account by following Section \ref{sec:pipeline-parallelism} and assuming either a 1F1B interleaving schedule \citep{narayanan2021efficient} or a ZB-H2 schedule \citep{qi2023zero}. For the 1F1B schedule, the bubble fraction \( f_b \) is computed following Equation \ref{eq:interleaved-pp-bubble-fraction}:
\begin{align*}
    f_b &= \frac{(\npp-1) + z}{(\npp-1) + z + im},\\
    z &= (i-1)\cdot\max(0, \npp - m),
\end{align*}
where \( i \) is the pipeline interleaving factor and \( m \) is the number of microbatches in the pipeline. For the ZB-H2 schedule, we assume \( f_b = 0 \) but require that the number of microbatches \( m \geq 2\npp - 1 \), as this is the condition required to achieve no pipeline bubble.

Data-parallel all-reduce communication time is assumed exempt from the pipeline bubble as this normally occurs in a separate phase.

\subsubsection{Scaling assumptions}
\label{sec:model-scaling-assumptions}

\begin{wrapfigure}{R}{0.5\textwidth}
\vspace{-35pt}
\centering
  \includegraphics[width=0.48\textwidth]{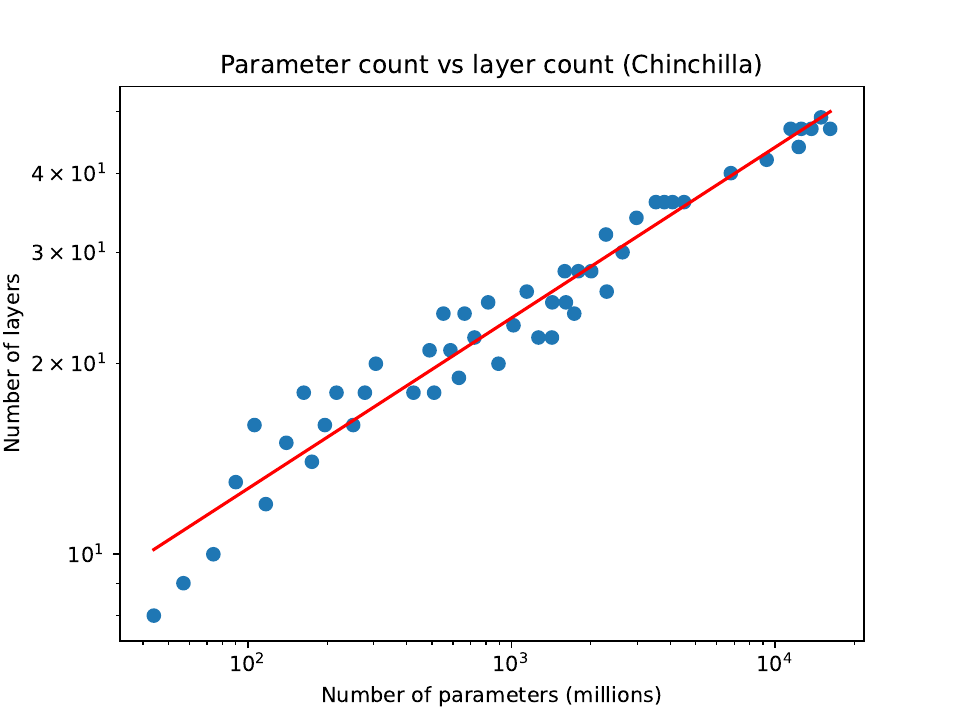}
  \caption{Parameter vs. layer count for models trained in \cite{hoffmann2022training}. The red line corresponding to the best power law fit is given by \( L = 3.67 \cdot (\nparams/10^6)^{0.27} \) and achieves an \( R^2 \) of \( 94.7 \% \).}
  \label{fig:layer-scaling-law}
\vspace{-80pt}
\end{wrapfigure}

We assume the following scaling relationship for the baseline scenario:

\begin{itemize}
    \item \textbf{Critical batch size:} \( b = 2^{22} \cdot E^{1/2} \cdot \left( \frac{T}{3 \cdot 10^{23} \flop} \right)^{1/6} \text{ tokens} \), where \( T \) is the training compute of a compute-optimal model
    \item \textbf{Number of MLP blocks:} \( L = 0.10056 \cdot (\dmodel \dff)^{0.3751} \)
    \item \textbf{Sparsity factor:} \( E = 8 \cdot \left( \frac{\dmodel \dff}{4 \cdot 12288^2} \right)^{1/2} \)
    \item \textbf{Feedforward dimension:} \( \dff = 4 \cdot \dmodel \)
    \item \textbf{Dataset size:} \( D = 20 \nparams \)
\end{itemize}

The scaling relation for the number of MLP blocks is informed by the scaling done in \cite{hoffmann2022training}. Figure \ref{fig:layer-scaling-law} shows the results we obtain by analyzing the information about layer scaling in the models trained in \cite{hoffmann2022training}, with the best-fitting scaling law \( L \approx 3.67 \cdot (\nparams/10^6)^{0.27} \) where \( L \) is the number of layers and \( \nparams \) is the number of model parameters.\footnote{If we directly substitute \( \nparams = 2 L \dmodel \dff \), these two scaling laws appear inconsistent with each other. The discrepancy is explained by the fact that the models trained by \cite{hoffmann2022training} have attention layers, while our toy model implying this expression for the parameter count only considers MLP blocks and ignores the QKV and post-attention projections, which in practice also contribute to the parameter count.}

These are not as well-justified as we would like them to be due to a lack of solid empirical evidence. We choose them to match cases where we have explicit information when we can (e.g. a sparsity factor of \( 8 \) for GPT-4, a batch size of \( 2^{22} \) tokens for GPT-3, \textit{et cetera}) and we base the scaling exponents on the discussions in relevant parts of Section \ref{sec:constraints}. Our framework allows for the input of alternative scaling assumptions.

\subsubsection{Time taken for a training run}
\label{sec:model-time-taken}

Putting together the considerations from Sections \ref{sec:model-matmul}, \ref{sec:model-network-communication} and \ref{sec:model-bubbles}, we compute the time taken per gradient step as

\begin{equation}
t_{\text{step}} = t_{\text{network latency}} + t_{\text{DP}, n} + \max \left( t_{\text{DP}, y}, \frac{\max(t_{\text{matmul}}, t_{\text{n-DP}, y}) + t_{\text{n-DP}, n}}{1 - f_b} \right)
\end{equation}

Here, the subscripts DP and n-DP of \( t \) refer to data parallel and non-data parallel communication time respectively, while the subscripts $ y $ and $ n $ (standing for "yes" and "no", respectively) refer to whether said communication can be overlapped with computation or other communication or not. In the Zero Bubble case only data-parallel communication counts toward $t_{\text{network latency}}$, as the rest can be hidden.

We combine this with the scaling relationships from \ref{sec:model-scaling-assumptions} to compute the time taken for the entire training run as
\[ t_{\text{run}} = n_{\text{steps}} \cdot t_{\text{step}} = \frac{D \cdot t_{\text{step}}}{b}, \]
where \( D \) is the training dataset size and \( b \) is the critical batch size. This is the final output of our model.

\subsection{Appendix: Effect of pipeline interleaving}
\label{sec:appendix-interleaving}

Suppose our model has $L = 12$ MLP blocks (numbered $0 \ldots 11$), and our pipeline has $\npp = 3$ stages (numbered $0 \ldots 2)$. In the non-interleaved ($i = 1$) case, then the MLP block $\to$ pipeline stage mapping is:
\begin{table}[h!]
\centering
\begin{tabular}{r|c|c|c|c|c|c|c|c|c|c|c|c|}
\textbf{MLP Block} & 0 & 1 & 2 & 3 & 4 & 5 & 6 & 7 & 8 & 9 & 10 & 11\\
\textbf{Pipeline Stage} & \multicolumn{4}{c|}{0} & \multicolumn{4}{c|}{1} & \multicolumn{4}{c|}{2}
\end{tabular}
\end{table}
\vspace{-5mm}

If instead we employ an interleaving factor $i$ (restricted so that $\npp \cdot i$ divides $L$), then on every forward and backward pass, each of the $m$ microbatches passes through the pipeline $i$ times rather than once, each time seeing only $L/i$ consecutive MLP blocks. On each pass through the pipeline, the $\npp$ pipeline stages are encountered in order, each responsible for $L/(\npp \cdot i)$ consecutive MLP blocks per pass.

So in the concrete example above, if we take $i = 2$, the mapping becomes:
\begin{table}[h!]
\centering
\begin{tabular}{r|c|c|c|c|c|c|c|c|c|c|c|c|}
\textbf{MLP Block} & 0 & 1 & 2 & 3 & 4 & 5 & 6 & 7 & 8 & 9 & 10 & 11\\
\textbf{Pipeline Stage} & \multicolumn{2}{c|}{0} & \multicolumn{2}{c|}{1} & \multicolumn{2}{c|}{2} & \multicolumn{2}{c|}{0} & \multicolumn{2}{c|}{1} & \multicolumn{2}{c|}{2}
\end{tabular}
\end{table}
\vspace{-5mm}

As there are now $\npp \cdot i$ ``virtual'' pipeline stages, the number of inter-stage interfaces has increased from $\npp - 1$ (in the interleaving-free case) to $\npp \cdot i - 1$. Therefore activations must be communicated approximately $i$ times as frequently. In exchange, devices spend less time idle in the pipeline bubble, as the pipeline fills up and clears out $i$ times as fast at the beginning and end of each batch.

Let us compute the bubble fraction precisely, which as usual is easiest from the perspective of the last stage. It spends $\npp - 1$ ``bubble'' steps waiting for that same number of previous stages before it sees its first microbatch, then sees all $m$ microbatches, for $m$ steps of work.

At this point, the microbatches are recycled through the pipeline for the next interleaving cycle. If $m < \npp$, then there are insufficient microbatches to fill the pipeline, and the last stage must wait $\npp - m$ steps for the first microbatch to reach the end again, at which point it has $m$ more steps of work. This recycling process repeats $i - 1$ times.

Tabulating, we have $\npp - 1$ initial bubble steps, $z := (i - 1) \cdot \max(0, \npp - m)$ inter-cycle bubble steps, and $im$ work steps, so our last stage's forward pass bubble fraction is
\begin{align}
    \label{eq:interleaved-pp-bubble-fraction}
    B_{\text{interleaved PP}} = \frac{\npp - 1 + z}{\npp - 1 + z + im}.
\end{align}
The situation looks the same in reverse on the backward pass. Furthermore, since all stages have the same amount of work to do, they must also have this same bubble fraction.

In the typical case when the number of microbatches $m$ can fill the pipeline depth $\npp$, then this is simply Equation \ref{eq:naive-pp-bubble-fraction}, with \( im \) substituted for \( m \). So we have multiplied the ``effective'' microbatch count by $i$, reducing the bubble accordingly, at the cost of increasing pipeline communication costs by roughly this factor as well.

\subsection{Appendix: Allowing for layer count and batch size scaling}
\label{sec:appendix-allowing-scaling-layer-batch}

\begin{figure}[h]
  \centering
  \begin{minipage}{0.5\textwidth}
    \centering
    \includegraphics[width=\textwidth]{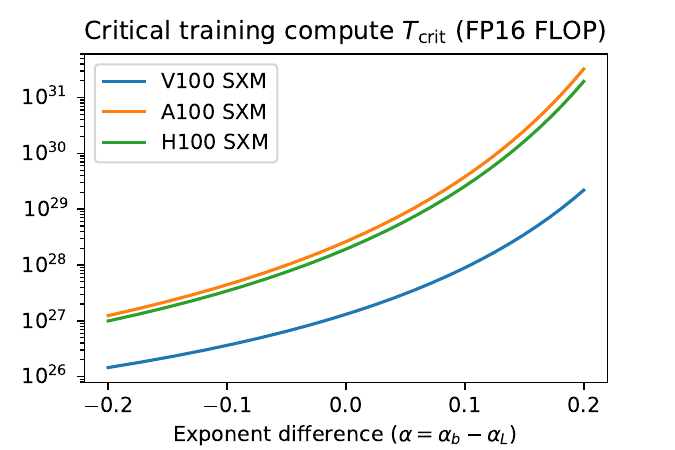}
  \end{minipage}%
  \hfill
  \begin{minipage}{0.5\textwidth}
    \centering
    \includegraphics[width=\textwidth]{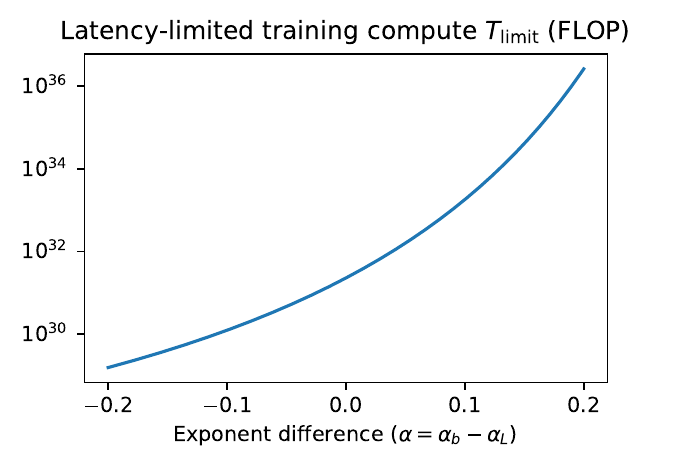}
  \end{minipage}
  \caption{The sensitivity of \( \tcrit \) and $\tlim$ for dense training runs, to the exponent difference \( \alpha_b - \alpha_L \), starting from a reference training run of $3 \times 10^{23} \flop$ for a model having 100 MLP blocks, using a batch size of 4M tokens. In Sections \ref{sec:utilization-cliff} and  \ref{sec:absolute-limit} we assume \( \alpha_b = \alpha_L \), in which case the effects of batch and layer scaling cancel out.}
  \label{fig:scaling-sensitivity}
\end{figure}

In Sections \ref{sec:utilization-cliff} and \ref{sec:absolute-limit} we assumed a ratio between the critical batch size $b$ and the number of MLP blocks $L$ that does not scale with the size of the training run, which we justified with some (weak) empirical evidence. However, our analysis can be generalized. Suppose instead the power law scaling relationships \( b = b_0 (T/T_0)^{\alpha_b}, \, L = L_0 (T/T_0)^{\alpha_L} \), where $T$ is the total training compute, and $b_0$, $L_0$ are the batch size and depth for a reference training run of $T_0$ compute. Then defining $\alpha := \alpha_b - \alpha_L$ and solving for $\tcrit$, the bandwidth bottleneck formula Eq. \ref{eq:ccrit} becomes
\begin{align*}
    \tcrit &= \Bigg[ \frac 1 {(960\mac) \cdot E} \left(\frac {b_0} {L_0} \cdot \frac 1 {{T_0}^\alpha} \cdot \frac{C \traintime} {\dcrit^2 \bcrit}\right)^2 \Bigg]^{\frac 1 {1 - 2\alpha}},
\end{align*}
while solving for $\tlim$, the latency limit formula Eq. \ref{eq:climit-latency-bottlenecked} becomes
\begin{align*}
    T_\text{limit} = \Bigg[ \frac {3\mac} {320 \cdot E} \left(\frac {b_0} {L_0} \cdot \frac 1 {{T_0}^\alpha} \cdot \frac{\traintime}{ t_L}\right)^2 \Bigg]^{\frac 1 {1 - 2\alpha}}.
\end{align*}
We plot the effects for dense training runs in Figure \ref{fig:scaling-sensitivity}, taking as our reference training run \( T_0 = 3 \times 10^{23} \) FLOP, with \( b_0 = 4 \times 10^6 \) tokens per batch and \( L_0 = 100 \) layers, assuming a $9 \mus$ lower bound on the matrix multiplication timescale, accounting for both intra- and inter-GPU latencies.

If correct, the results in \cite{deepseekai2024deepseek} imply an aggressive \( \alpha_b = 0.3271 \), yielding \( \alpha = \alpha_b - \alpha_L \approx 0.2 \). The impact is about three orders of magnitude of training compute, highlighting the importance of better understanding the scaling of the optimal batch size and model depth. At this scale, the relevant constraint becomes economic: acquiring a sufficiently large cluster and the energy to operate it.


\end{document}